\documentclass[pre]{revtex4}
\usepackage{graphicx}
\usepackage{amsmath}
\usepackage{bbold}
\usepackage{amssymb}
\usepackage{color}
\usepackage[normalem]{ulem}
\usepackage{epsfig}
\usepackage{amssymb,amsfonts,amsmath, color}
\usepackage{array}
\usepackage{makecell}
\usepackage{float}
\renewcommand{\arraystretch}{2}

\begin{document}

\title{Numerical Study of Interaction Network Structures in Competitive Ecosystems} 

\author{David A. Kessler and Nadav M. Shnerb}

\affiliation{Department of Physics, Bar-Ilan University,
Ramat-Gan IL52900, Israel.}

\begin{abstract}
We present a numerical analysis of local community assembly through weak migration from a regional species pool. At equilibrium, the local community consists of a subset ("clique") of species from the regional community. Our analysis reveals that the interaction networks of these cliques exhibit nontrivial architectures. Specifically, we demonstrate the pronounced nested structure of the clique interaction matrix in the case of symmetric interactions and the hyperuniform structure seen in asymmetric communities. 
\end{abstract}

\maketitle
\section{Introduction}

Diverse and even highly diverse assemblages of biological types, such as multiple species within a local community or various genetic types within a population, are ubiquitous in nature~\cite{chesson2000mechanisms,levine2017beyond}. Prominent examples include trees in tropical forests~\cite{ter2013hyperdominance,maritan1}, coral reefs~\cite{connolly2014commonness}, freshwater plankton~\cite{stomp2011large}, and others. More recently, communities of human and soil microorganisms have garnered significant attention~\cite{friedman2017community,bashan2016universality,yonatan2022complexity}. Understanding the mechanisms that enable coexistence and the drivers that shape community structure remains a challenging and practically important puzzle.

In the past decade, significant progress has been made in mapping the range of coexistence possibilities by examining the dynamics of a local community (an 'island') subject to weak migration from a regional pool (a 'mainland')~\cite{fisher2013niche,kessler2015generalized,bunin2017ecological,barbier2018generic}. When the regional pool is rich and the immigration rate is negligible, the species on the island can be divided into two groups: one group consists of species with extremely low abundance, proportional to the immigration rate from the mainland, while the other group consists of species with much higher abundances, forming a ``clique'' of resident species. 

In some cases, the clique of resident species is stable and uninvadable, meaning that it represents a steady-state solution of the corresponding deterministic equation, and no other species from the mainland can successfully invade from rarity~\cite{fried2016communities,fried2017alternative}. In other cases the dynamics does allow invasion, but the species turnover time is still relatively long when the immigration rate is slow~\cite{arnoulx2024many,mallmin2024chaotic}.    

Therefore, even if the interactions between species in the regional community are random in nature, this is not true for the interactions between species within the clique. Each of these cliques reflects a single choice from an immense number of possibilities (factorial in the number of mainland species). Both the competitive exclusion principle~\cite{tilman1982resource} and May's analysis of the complexity-diversity problem~\cite{may1972will}  suggest that, in diverse systems, stable coexistence is more likely when interspecies interactions are weak and their variability is low. Consequently, it is expected that the species dynamically selected to form a stable, uninvadable community (or, at least,  a relatively long-lived clique) will be characterized by precisely these features -- namely, weaker and more similar interactions compared to the average interactions among all species on the mainland.

Our numerical findings indicate that this is not enough. The observed values of these summary statistics parameters (mean and variance of the interaction matrix) are not sufficient to stabilize the clique. In particular, reshuffling the elements of the interaction matrix within the clique causes it to lose its stability, meaning that the stability is built not only on the summary statistics of the matrix elements but also on the specific structure of the interaction network between the species.

The aim of this paper is to demonstrate a signal feature of the interaction network structure for each of two cases that can be considered prototypes of competitive dynamics. The first case is that of a symmetric interaction matrix, where for each pair of species, the pressure exerted by the first on the second is equal to the pressure exerted by the second on the first. The second case is that of an asymmetric interaction matrix, where there is no correlation between the pressure exerted by species A on species B and the pressure exerted by species B on species A. We focus on competitive interactions, neglecting mutualism or predation. 

Both cases (symmetric and asymmetric) may describe competition for common resources. In the symmetric case, the yield of species consuming the same resource is identical. The asymmetric case corresponds to uncorrelated yields. It is typically assumed that the interactions in competitive communities one finds in nature lie somewhere between these boundaries.

Our numerical experiments suggest that, in the symmetric case, the interaction matrix of the assembled clique is nested, where, for each species, the number of strong competitors is inversely proportional to its rank abundance. In the asymmetric case the nested structure is less pronounced, but the interaction matrix admits a new feature, hyperuniformity. Both concepts - nestedness and hyperuniformity -   will be explained, quantified, and examined below.

This paper is organized as follows. In the next section, we provide a general background to the problem, describing the possible dynamics of the island clique and the scaling parameters used in the analysis. In Section \ref{sec3}, we demonstrate that the summary statistics of the emerging cliques cannot fully explain their stability. In Section \ref{sec4}, we explore the different architectures of the clique networks, specifically their nestedness and hyperuniformity. Finally, we discuss the implications of our findings.

\section{Model, scaling parameter, phases and bifurcations} \label{sec2}

\citet{barbier2018generic}  demonstrated that a wide range of ecological community models yield the same qualitative phase diagram. Therefore, we use the standard workhorse of the field, the generalized Lotka-Volterra model, with standard parameters related to interaction strength,  heterogeneity, and symmetry. Specifically, we analyze the dynamics of,
\begin{equation}\label{eq1}
\frac{dN_i}{dt} = N_i \left( 1-N_i -\sum_{j \neq i}^S \alpha_{i,j} N_j \right) + \lambda_i.
\end{equation}
Here $\alpha_{i,j} = \alpha_0  + \eta_{i,j} $ is the interaction matrix and the total number of species in the regional pool is $S$. The $\eta_{i,j}$-s  are picked from a zero mean normal distribution whose variance is $\sigma^2$, so the relevnt parameters of the model (\emph{\`a la} \cite{kessler2015generalized,barbier2018generic}) are $\alpha_0$ and $\sigma$.  This matrix is symmetric if $\eta_{i,j} = \eta_{j,i}$ and asymmetric if  $\eta_{i,j}$ and $\eta_{j,i}$ are picked in an uncorrelated manner. As usual, we consider the case in which $\lambda_i = \lambda\ll 1$ for all species (equal and tiny migration rates). Throughout this work we set $\lambda = 10^{-10}$.

In our simulations an initial condition  $N_i(t=0)$ is picked for every species from a uniform distribution between zero and one. The set of couples ODEs (\ref{eq1}) is then integrated using Matlab's ode23s routine until $t = 2 \cdot 10^5$, whereupon the clique species are identified and the structure of the matrix is analyzed. In some simulations, we turn off the immigration at $t=2 \cdot 10^5$, setting $\lambda=0$ at that point, and continue the simulation out to $t=3\cdot 10^5$.

At the end of the simulation, we identify the clique of island species, which are all the species whose abundance is larger than $\sqrt{\lambda}$. A fairly good test for stability (see Appendix \ref{apA}) is that  the abundances of the clique species satisfies
\begin{equation}
    {\cal Q} {\cal N} =   \mathbb{1}, 
\end{equation}
where $ {\cal Q}$ is the interaction matrix of the clique, ${\cal N}$ is the vector of abundances of clique species and $\mathbb{1}$ is the  vector  of all $1$s, of length the size of the clique. 

Our main focus here is on identifying nontrivial structural features
 of the clique matrix $ {\cal Q}$. In particular, we are interested in how these features vary as a function of the two summary statistics parameters,  $\alpha_R$, the mean over all non-diagonal entries, and $\sigma_R$, the variance of this set.

\subsection{The symmetric case}

\begin{figure}
\begin{center}
\includegraphics[width=0.45\textwidth]{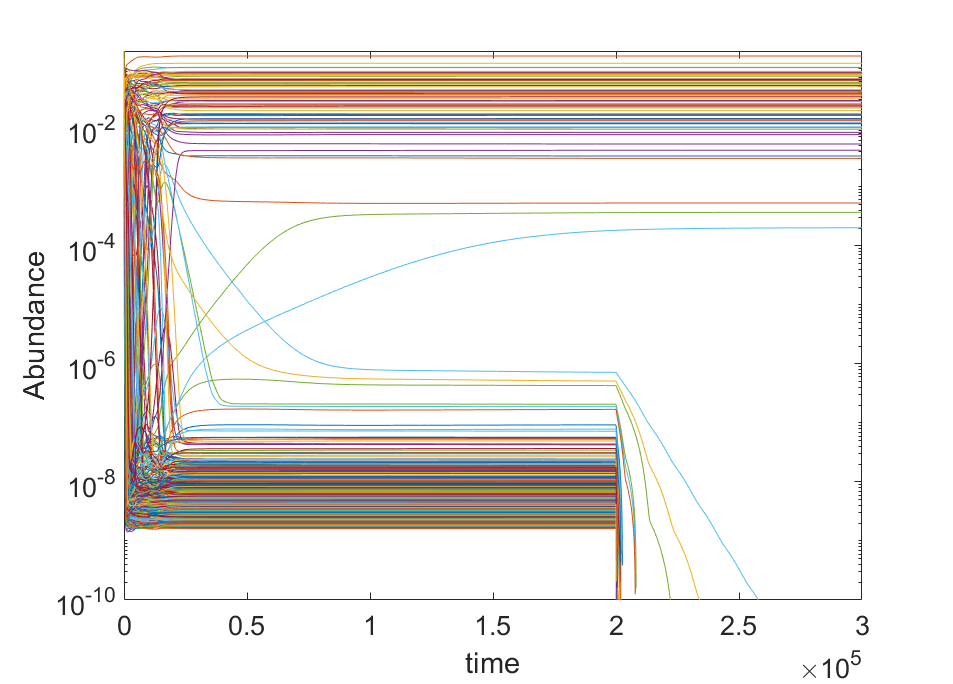}
\end{center}
\vspace{-0.5 cm}
\caption{{\bf Dynamics, Symmetric Case: } A typical dynamics of a diverse community, when the interaction matrix is symmetric, as obtained from a numerical integration of  Eq. (\ref{eq1}). Each curve shows the abundance of a single species vs. time. The total number of species (the diversity of the regional community) is  $S = 500$. Interaction matrix elements $\alpha_{i,j}$ were picked from a normal distribution with mean $\alpha = 0.4$ and standard deviation $\sigma = 0.04$. Since the interaction matrix is symmetric, $\alpha_{i,j} = \alpha_{j,i}$. Immigration from the regional pool is, $\lambda = 10^{-10}$, for $0 \le t \le 2 \times 10^{5}$. During this period, the community splits into two groups of species. The abundance of resident species is ${\cal O}(1)$, while the abundance of passenger species, that owe their existence to migration from the regional pool, is ${\cal O}(\lambda)$. The clique of $Q=65$ resident species is uninvadable, as none of the $S-Q$ passenger species may invade.  At $t=2 \times 10^{5}$  the connection to the regional pool is lost, i.e., $\lambda$ is set to zero.  the passenger species went extinct while the residents are shown to persist as a feasible and stable clique.   \label{fig1}}
\end{figure}

For the symmetric case, Figure \ref{fig1} shows the typical dynamics of the system. After some transient time, the local community settles into a stable solution where species are divided into two groups:
$Q$ resident species with abundances of ${\cal O}(1)$, independent of $\lambda$,  and $S-Q$ passenger species whose presence on the island depends on migration, resulting in abundances on the order of ${\cal O}(\lambda)$.  There is typically a gap between the abundances of these two groups, which is centered roughly at $N \sim \sqrt{\lambda}$. These results hold as long as $\lambda$ is much smaller than any other quantity in the system but still non-zero.

Figure \ref{fig1} also indicates that the clique of resident species is not only uninvadable (meaning that low-abundance species do not grow) but also stable by itself. Once the system reaches its steady state, migration is needed only to support the passengers. The sub-community of $Q$ resident species is feasible and stable even after migration is turned off, and all passenger species go toward extinction.

\begin{figure}
	%\vspace{-3.cm}
	\centering{
		\includegraphics[width=7cm]{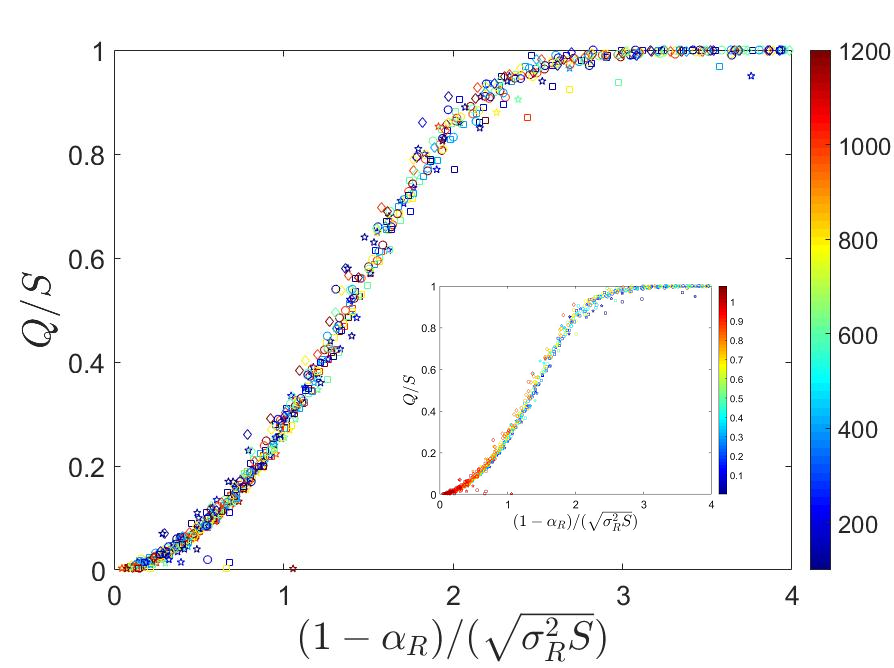}} \includegraphics[width=7.5cm]{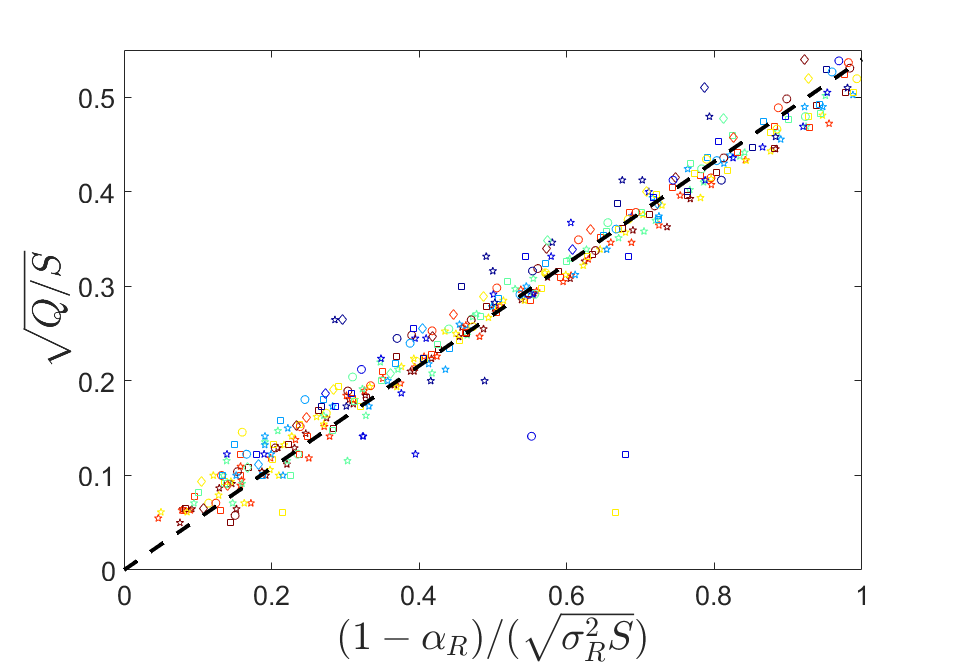}
	%\vspace{-3.cm}
\caption{{\bf Clique Size, Symmetric Case: } Left panel, main: the ratio between clique size $Q$ and the number of species in the regional pool $S$,  plotted against the scaling parameter $X = (1-\alpha_R)/\sqrt{\sigma_R^2 S}$.  Each point reflects the result of a single run. Markers correspond to different values of $\sigma$: $0.005$ (diamonds), $0.01$ (circles), $0.02$ (squares) and $0.04$ (pentagons). Color (see color bar) indicates the value of $S \in [100..1200]$. The same datasets are presented in the inset, now color indicates different values of $\alpha \in [0..1.1]$. The data collapse is evident in both cases. Right panel: the small-$X$ sector of the left panel, where Eq. (\ref{eq5}) holds. Here the \emph{square root} of $Q/S$ is plotted against  $x$, showing that the behavior of ${\cal F}(X)$ as $X \to 0$ is quadratic in $X$. The gap at the vicinity of zero was predicted by Song et al.~\cite{song2019consequences}. The dashed line corresponds to $\sqrt{{\cal F}(x)}  = 0.54 x$.   \label{fig2}}
\end{figure}

As suggested in~\cite{bunin2017ecological}, the expected ratio between the diversity of the resident clique, $Q$ and the mainland diversity $S$, is a function of a single parameter,
\begin{equation}
X \equiv \frac{1-\alpha_R}{\sqrt{\sigma_R^2 S}}
\end{equation}
so that
\begin{equation} \label{eq4}
\frac{Q(\alpha_R,\sigma_R,S)}{S} = {\cal F}\left(X\right).
\end{equation}
This general scaling relationship is demonstrated in Figure \ref{fig2}.  In the regime of small $X$,  ${\cal F}(X) \sim X^2$ (Fig \ref{fig2}, right panel). Accordingly, as $\alpha_R \to 1$ (strong competition) or $\sigma_R^2 S \to \infty$ (heterogeneous and diverse community)  the clique size $Q$ is independent of the number of species in the regional pool $S$ and satisfies,
\begin{equation}\label{eq5}
Q = \left(\frac{1-\alpha_R}{\sigma_R}\right)^2.
\end{equation}

The parameter $X$ is proportional to the ratio between the "width" of the eigenvalues cloud for a random Gaussian matrix of size $S$ and the location of its center~\cite{allesina2015stability}. May's stability criterion requires $X$ to be smaller than some constant of order one~\cite{allesina2012stability}. When Eq. (\ref{eq5}) holds, i.e., when $Q$ is $S$ independent, there are many alternative combinations of feasible and uninvadable $Q$-subcommunities, each of which is marginally stable (one Lyapunov exponent almost touches zero). When $X$ is larger, the local community displays a single stable equilibrium~\cite{bunin2017ecological}.

Technically speaking, this system supports two or three types of bifurcations as the parameter $X$ decreases. For very large $X$ all $S$ species coexist ($Q=S$) in a unique equilibrium. As $X$ decreases below $\sqrt{2 \ln S}$ some species are lost~\cite{bizeul2021positive} through a series of transcritical bifurcation, which at finite $\lambda$ becomes imperfect. The transition to a marginally-stable, multiple attractors state takes place via a series of saddle-node bifurcations~\cite{bunin2017ecological}. The deterministic dynamics eventually reaches a stable state, but noise may cause transitions between states and intermittent behavior~\cite{kessler2015generalized}, as demonstrated in recent experiment~\cite{hu2022emergent}.

\subsection{The asymmetric case} 

\begin{figure}[h]
	%\vspace{-3.cm}
	\centering{
		\includegraphics[width=5cm]{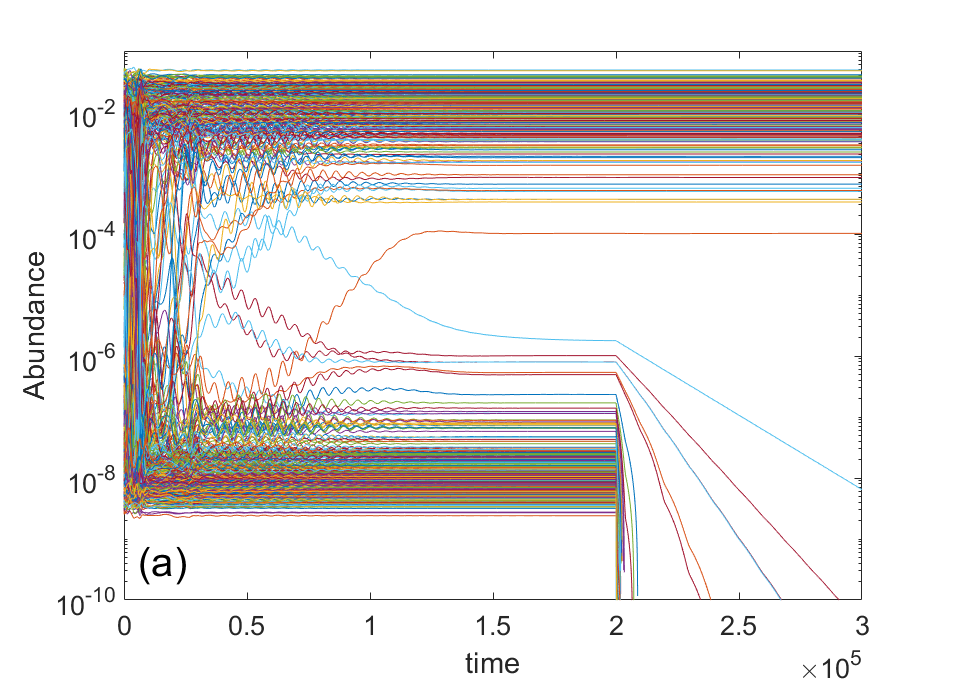} \includegraphics[width=5cm]{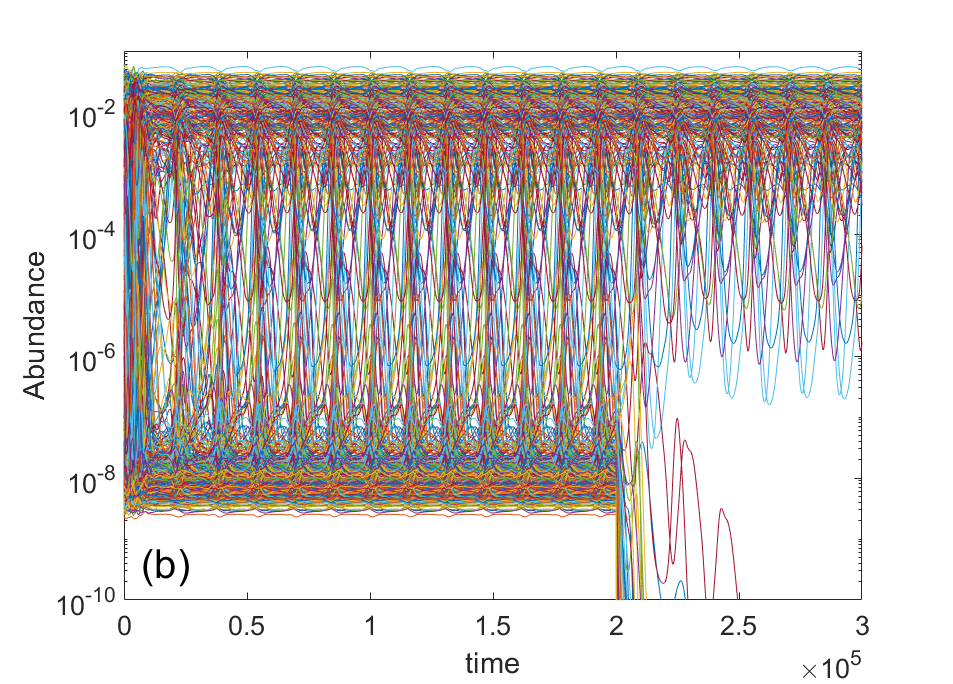} \includegraphics[width=5cm]{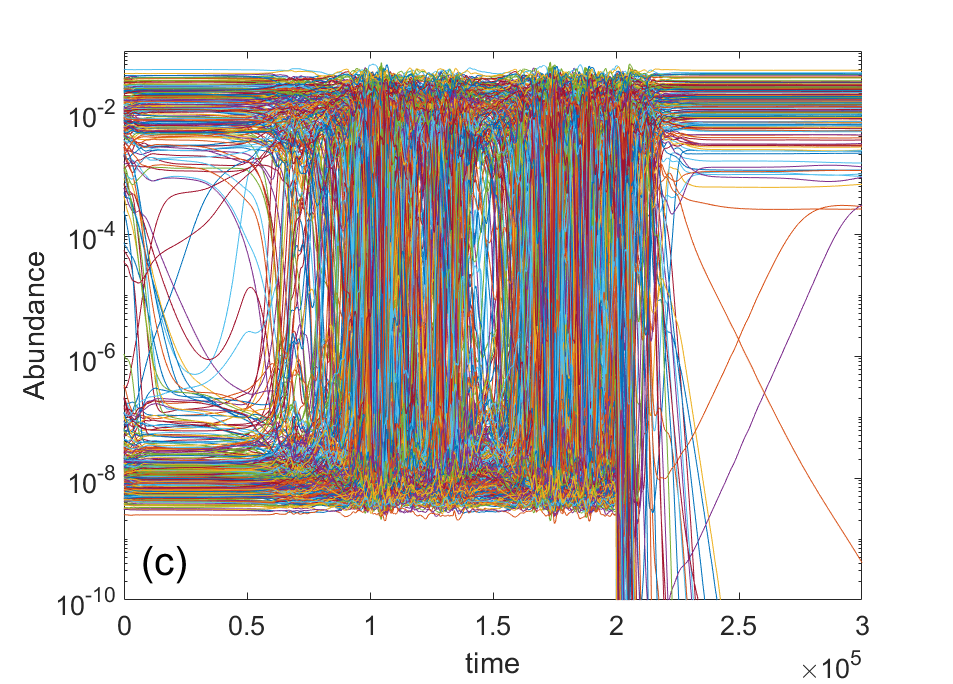}
	%\vspace{-3.cm}
	\caption{{ \bf Dynamics, Asymmetric Case: } Panel (a-c): Hopf bifurcation and the transition to chaos, typical characteristics of the non-symmetric interaction matrix. Shown here are the results of numerical integration of Eq. (\ref{eq1}), where the interaction matrix is asymmetric, namely, $\alpha_{i,j}$ and $\alpha_{j,i}$ were picked independently from a normal distribution with $\sigma = 0.05$, and the number of species in the regional community is, as in Fig. \ref{fig1}, $S=500$.  Right below the Hopf bifurcation ($\alpha = 0.35$, panel a) the system shows slowly decaying oscillations. Right above the bifurcation, it admits stable oscillations ($\alpha = 0.355$, panel b). At higher values of $\alpha$  ($0.38$, panel c) chaos becomes global, with trajectories that cross through the passenger-resident gap. All three panels correspond to the same interaction matrix (same set of $\eta_{i,j}$ values) and differ only by a shift in $\alpha_0$.  As immigration halts ($\lambda =0$ at $t>2 \times 10^5$) the passenger population collapses, leaving a stable resident community (panel a), or a stable limit cycle of resident species (panel b). The behavior of a globally chaotic community in that case is more subtle and involves aging, see \cite{arnoulx2024many}.   \label{fig3}}}
\end{figure}

Although we are dealing with a community where all interactions are competitive, when the interactions are asymmetric there can be cases of 'rock-paper-scissors': in each interaction between two species, one suffers more and the other suffers less, so there may be non-transitive relationships where A wins against B in a pairwise competition, B wins against  C, and C wins against A. As a result, the system may support Hopf bifurcation, above which the dynamics becomes periodic (limit cycle) or chaotic (see Figure \ref{fig3}).  Even in the periodic and the chaotic phases, there is still a distinction between the clique of species with abundance larger than $\sqrt{\lambda}$ and those with smaller abundance. However, the clique is not uninvadable and species turnover appears.

\begin{figure}
	%\vspace{-3.cm}
	\centering{
		\includegraphics[width=0.45\textwidth]{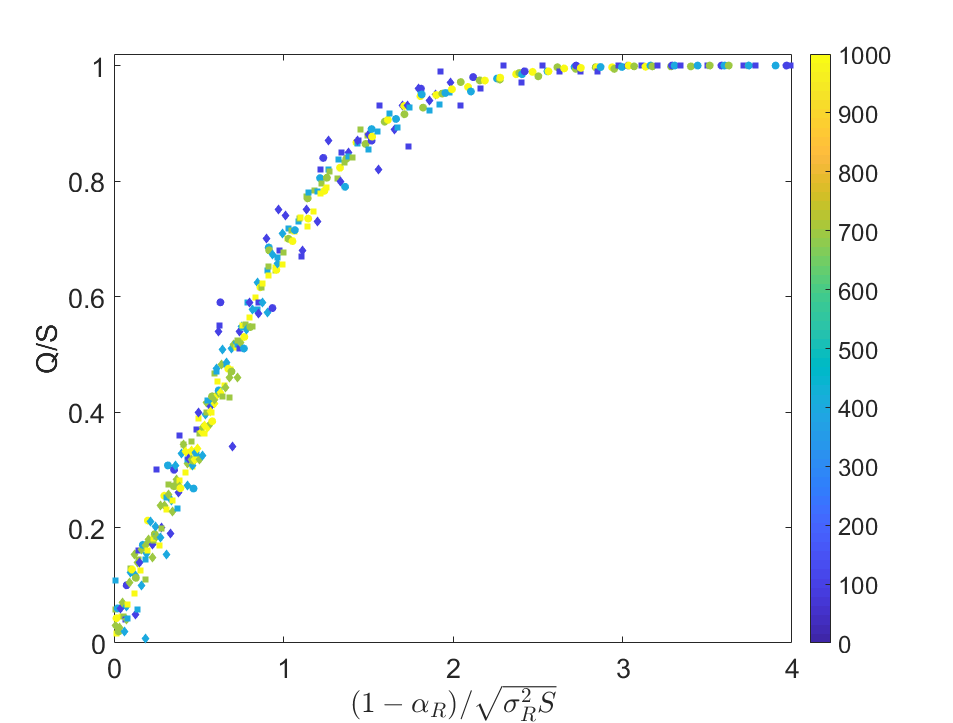}}
	%\vspace{-3.cm}
	\caption{{\bf Clique Size, Asymmetric Case: }The ratio between clique size $Q$ and the number of species in the regional pool $S$,  plotted against the scaling parameter $X = (1-\alpha_R)/\sqrt{\sigma_R^2 S}$ for an asymmetric interaction matrix.  Each point reflects the result of a single run. Markers correspond to different values of $\sigma$: $0.005$ (diamonds), $0.01$ (circles), $0.02$ (squares) and $0.04$ (pentagons). Color (see color bar) indicates the value of $S \in [100..1000]$.  Note that ${\cal F}(x)$ grows linearly with $X$ for small $X$ values, suggesting that the number of abundant species, $Q$, scales with $\sqrt{S}$ in this regime.  \label{fig3a}}
\end{figure}

Still, data collapse can be achieved by plotting the ratio between the clique size $Q$ and the size of the regional community $S$ as a function of the scaling parameter $X$, as seen in Figure \ref{fig3a}. However, when $X$ is small, the cliques in question (all species with abundance above the square root of lambda) are unstable and subject to invasions. This suggests that the cliques themselves have a certain structure that we will attempt to trace further.

\section{What makes a clique stable? } \label{sec3}

As we have seen, often (though not always), the local community in a competitive system of many species is stable and resistant to invasion. The question that arises is: what property grants it these features?

The immediate suspicion falls on the summary statistics of the clique interaction matrix, namely the average and heterogeneity of the elements of this matrix. If the interactions within the clique are weaker or more homogeneous, the clique is more likely to be stable. A second possibility is that the mean and variance of the clique matrix are almost identical to those of the regional community, yet the clique persists because it contains fewer species. A third possibility is that the clique possesses certain structural properties, meaning it cannot be characterized solely by the mean and variance of the competition matrix elements.

\begin{figure}
	%\vspace{-3.cm}
	\centering{
		\includegraphics[width=0.45\textwidth]{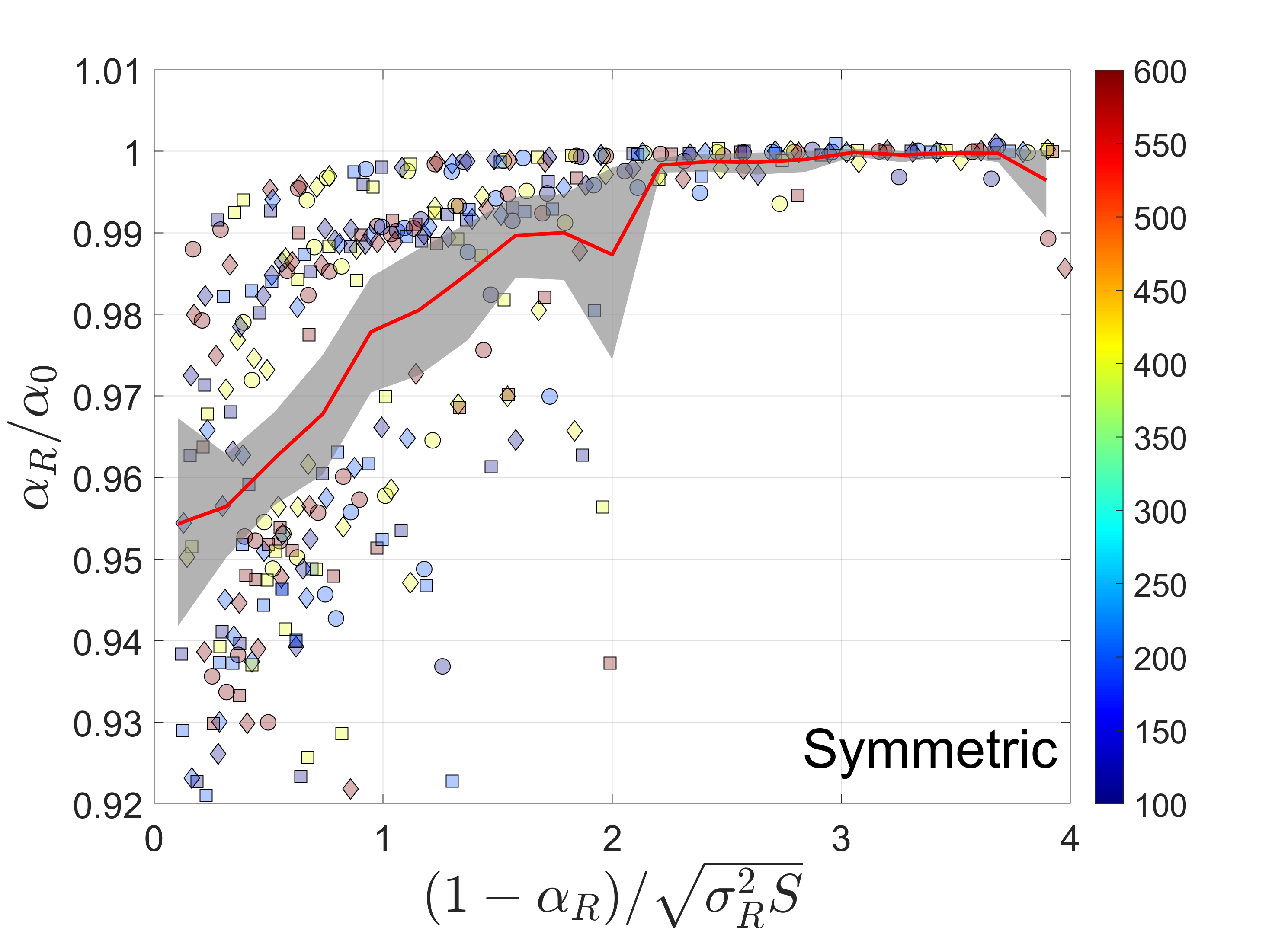} \includegraphics[width=0.45\textwidth]{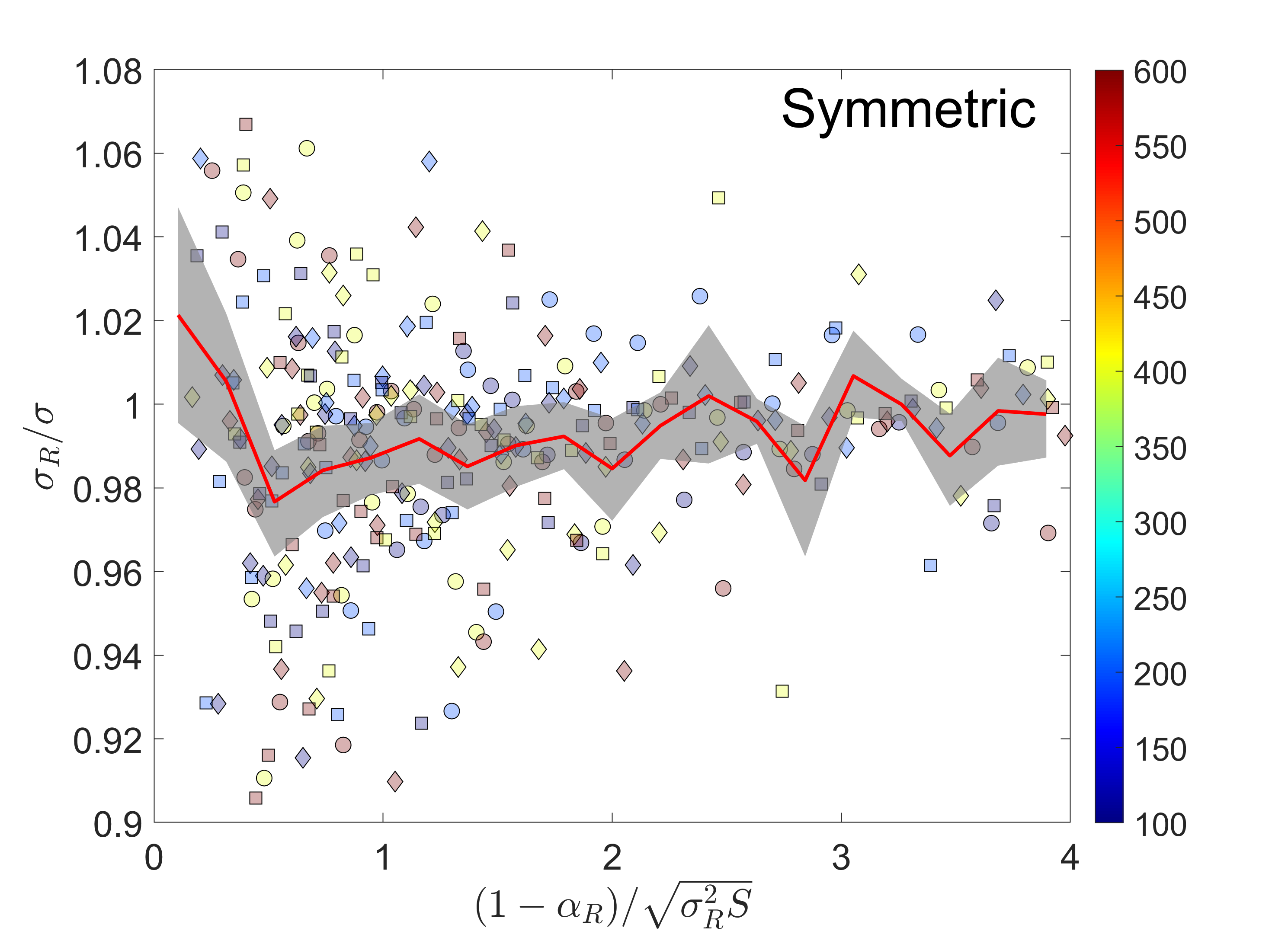} \\ \includegraphics[width=0.45\textwidth]{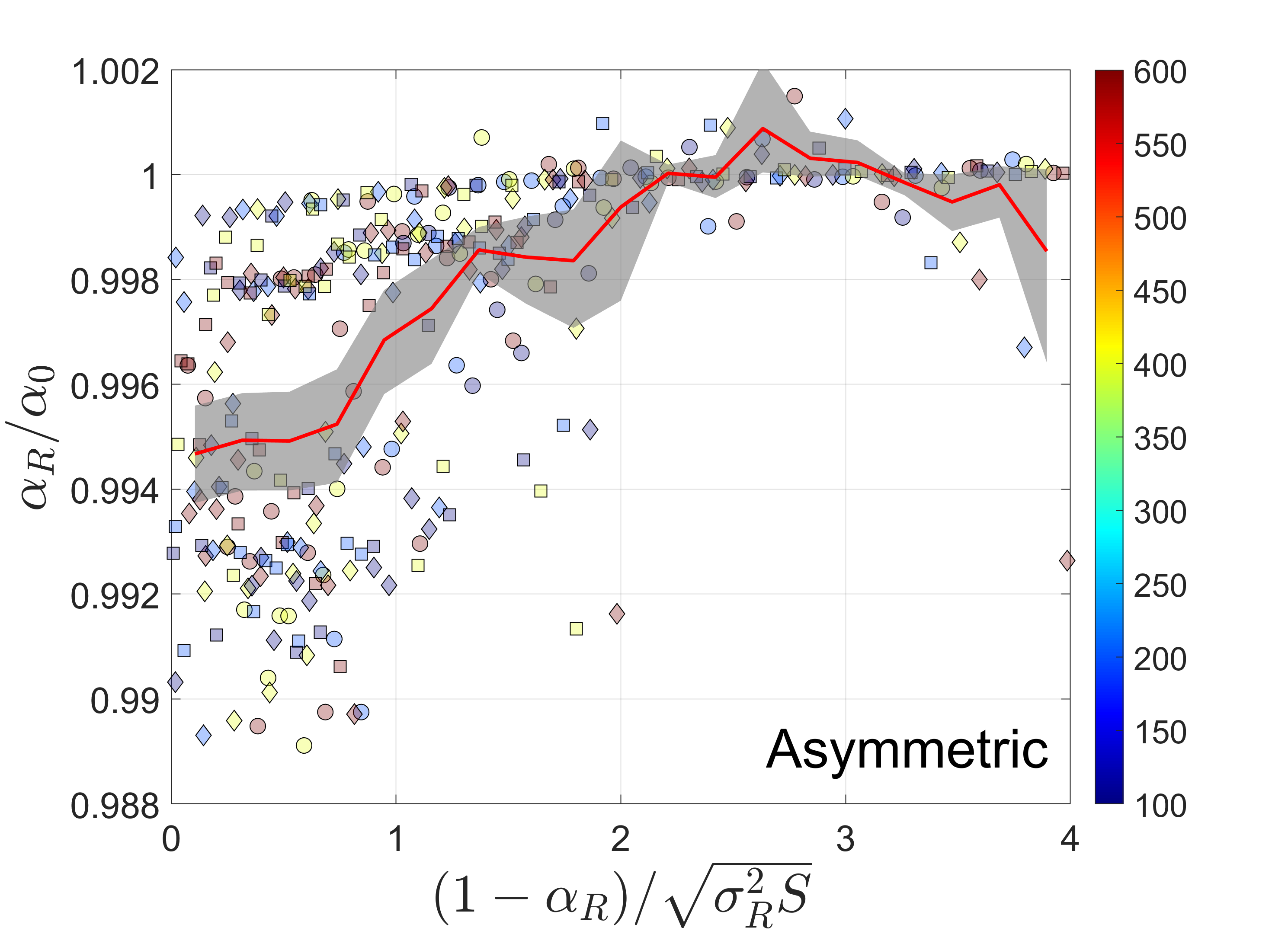} \includegraphics[width=0.45\textwidth]{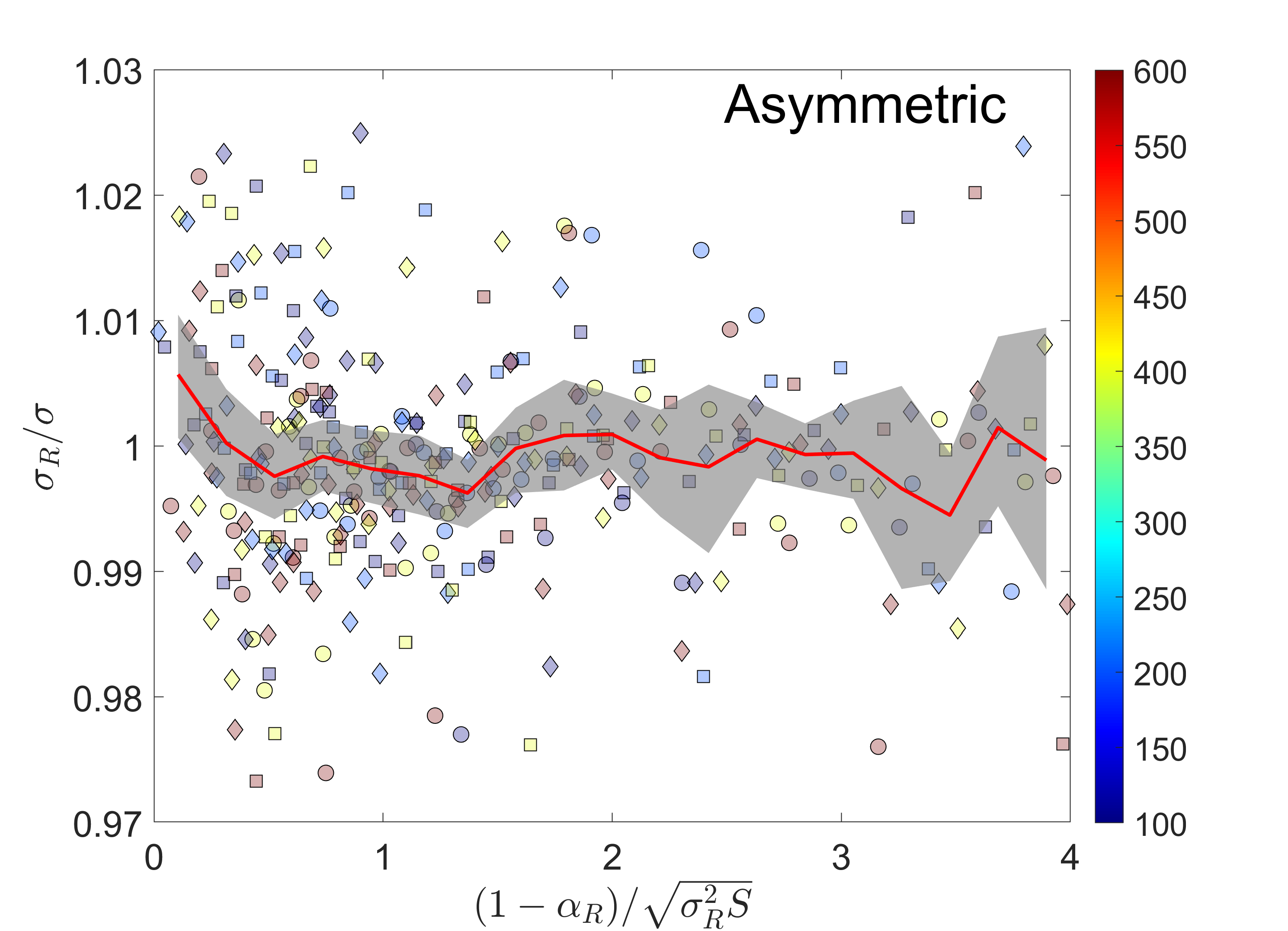} }
	%\vspace{-3.cm}
	\caption{{\bf Renormalized Parameters: } The renormalized values of $\alpha$ and $\sigma$ vs. the scaling parameter $X$, for symmetric (upper panels) and asymmetric (lower panels) communities.  In both cases, the mean values of $\sigma_R$ for the cliques are almost the same as its mean in the regional community (right panels). The mean value of $\alpha_R$ drops by about $5 \%$ in symmetric communities in the strong competition limit, and much less in the asymmetric case. Each marker represents the outcome of a single simulation run, with color indicating $S$ values and shape representing $\sigma$ (circle for $0.01$, square for $0.025$, and diamond for $0.05$). The solid red line shows the moving average, and the gray shaded area represents the $95 \%$ confidence interval. \label{fig3b}}
\end{figure}

Let us first quantify the summary statistic parameters. In the regional community, the mean over all  $\alpha_{i,j}$s is, by definition, $\alpha_0$, and the standard deviation is $\sigma$. As explained, the corresponding parameters for the clique matrix  ${\cal Q}$ and   $\alpha_R$ and $\sigma_R$.  The ratios $\alpha_R/\alpha_0$ and $\sigma_R/\sigma$ are presented in Figure \ref{fig3b}. Generally, these parameters are close to unity, indicating no significant deviations between the summary statistics parameters of the local and the regional community. 

To quantify this, and to test the role of the clique size in the clique feasibility (and presumably its stability),   we took {\it only} the stable cliques (see Appendix \ref{apA}) and permuted the elements of their matrices twenty times. We then solved, for each permutation, the linear problem $  {\cal Q}_p {\cal N} =   \mathbb{1}$, where ${\cal Q}_p$ is the permuted interaction matrix of the clique. The permuted local community is feasible if it produced only positive values for all $N$s. Each stable clique was assigned a score between one (all permutations are feasible) and zero (none of the permutations are feasible).

As shown in Figure \ref{fig3c}, once we reach the range of medium to strong interactions (medium-small $X$), most of the permuted cliques are not stable. This indicates that the summary parameters are not sufficient to explain clique stability, at least for medium or stronger coupling. As suggested earlier, additional structural factors are evidently  necessary, which we will address in the following section.

\begin{figure}
	%\vspace{-3.cm}
	\centering{
		\includegraphics[width=0.45\textwidth]{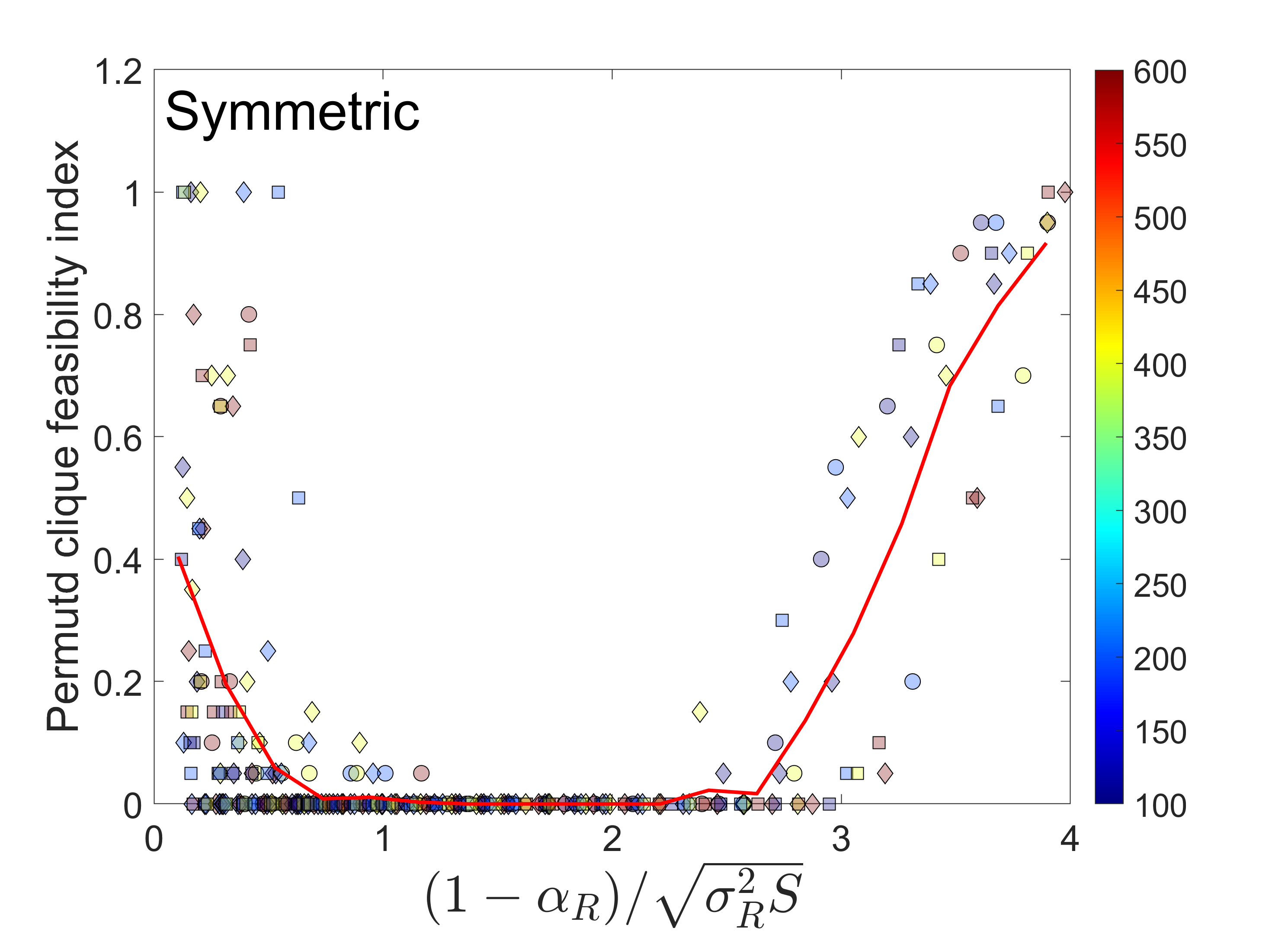}\includegraphics[width=0.45\textwidth]{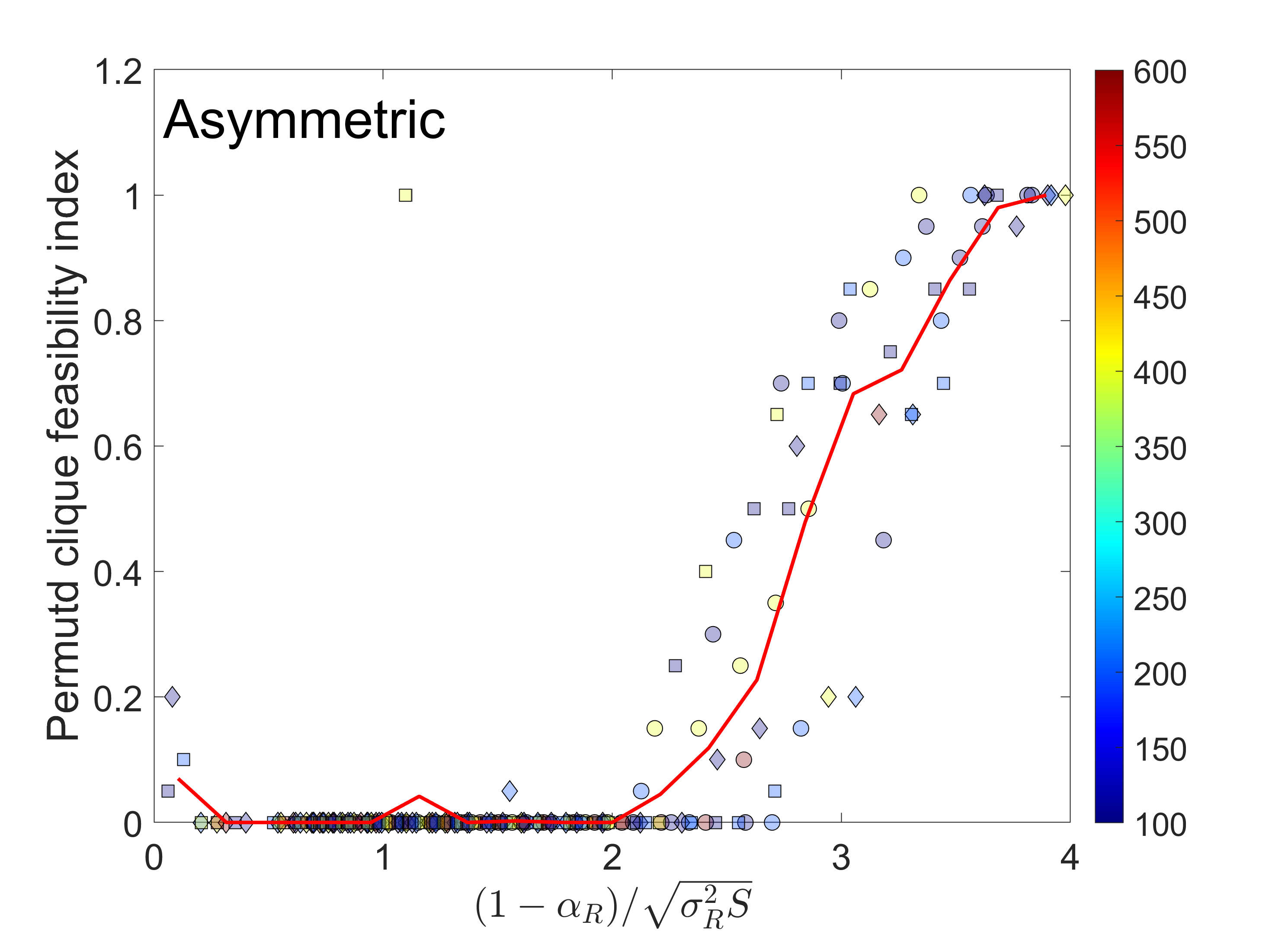}  }
	%\vspace{-3.cm}
	\caption{{\bf Feasibility Index: } The feasibility index  of the clique interaction matrix. For every {\it stable} clique we check the feasibility of 20 independent permutations of its interaction matrix ${\cal Q}$. We then solve   $  {\cal Q}_p {\cal N} =   \mathbb{1}$, where $  {\cal Q}_p$ is the permuted  interaction matrix and check for feasibility (${\cal N}_i>0$ for all $i$s).  The feasibility index is zero if all permutations are unfeasible, and one when all permutations are feasible. As shown in the graphs, the feasibility index decreases sharply to zero for intermediate values of the interaction parameter, and although we see a few high points at small values of $X$, the bulk of the distribution actually remains close to zero even there. The solid red line represents the moving average and the other color and marker codes are identical to those in Fig \ref{fig3b}.   \label{fig3c}}
\end{figure}

\section{Network Structure: Nestedness and hyperuniformity} \label{sec4}

In this section, we present our main results regarding the emergent structure of the clique interaction matrix, namely nested for symmetric interactions and hyperuniform for asymmetric interactions. We briefly explain the concepts of nestedness and hyperuniformity, describe the metrics used to quantify these properties, and examine their relevance to the clique interaction matrix in both symmetric and asymmetric cases.

\subsection{Nestedness}

An example of a  perfectly {\bf nested}  interaction matrix for a $5$ species community is given by
\[
{\cal Q} = \left( \renewcommand{\arraystretch}{1} % Adjust row height
\begin{array}{ccccc}
1 & 0 & 0 & 0 & 0 \\
0 & 1 & 0 & 0 & .9 \\
0 & 0 & 1 & .9 & .9 \\
0 & 0 & .9 & 1 & .9 \\
0 & .9 & .9 & .9 & 1
\end{array} \right)
\]
In this matrix, the competition of each species with itself (the diagonal terms) was set to unity. It can be observed that the competitors (including itself) of the species with index $n$ row species are a subset of the competitors of the species with index $n+1$.  

The identification of nested structure is often subject to debate~\cite{beckett2014falcon}, as the order of species (hence the order of rows and columns in the interaction matrix) can be arbitrarily changed, and statistically, some of these arrangements are likely to exhibit a more nested structure. Here, we do not face this issue because the order of species we use is always from most common to rare. The first row of the interactions matrix corresponds to the most abundant species, the second row the second most abundant and so on. 

\begin{figure}
	%\vspace{-3.cm}
	\centering{
		\includegraphics[width=5cm]{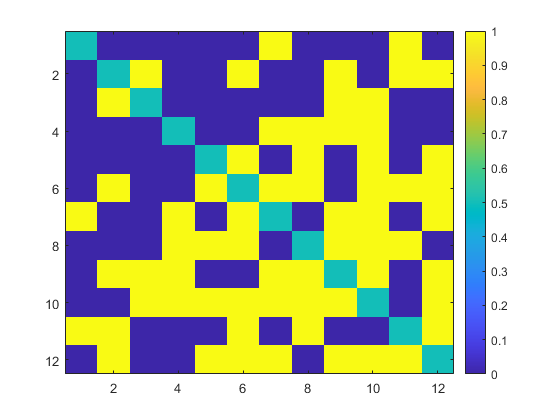} \includegraphics[width=5cm]{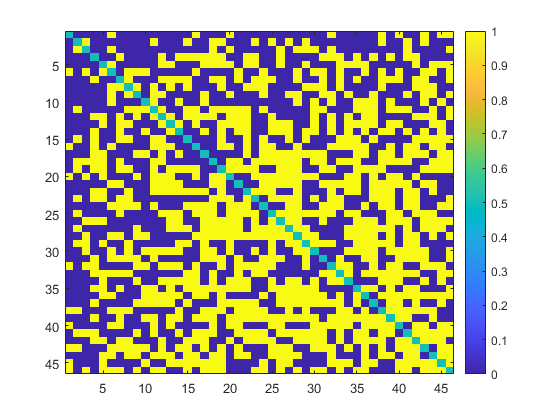}\includegraphics[width=5cm]{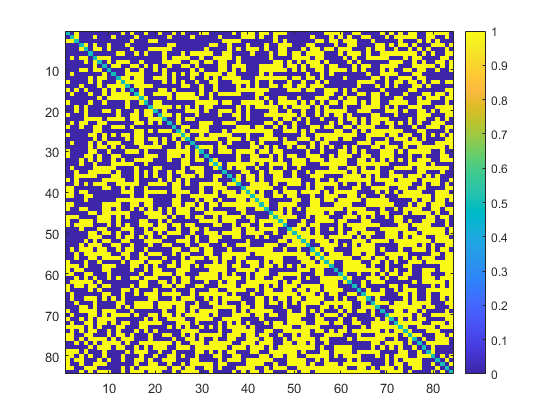}\\
\includegraphics[width=5cm]{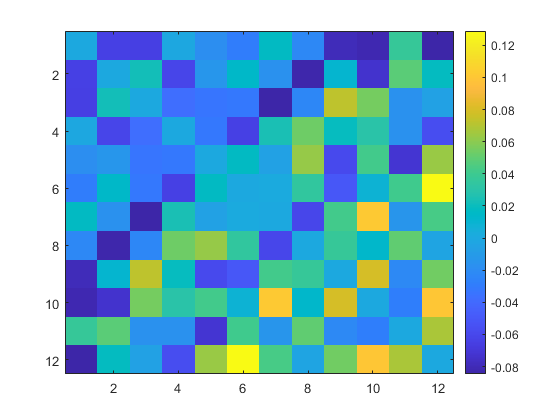} \includegraphics[width=5cm]{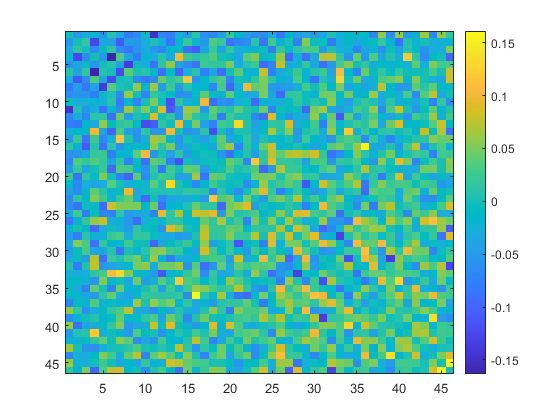}\includegraphics[width=5cm]{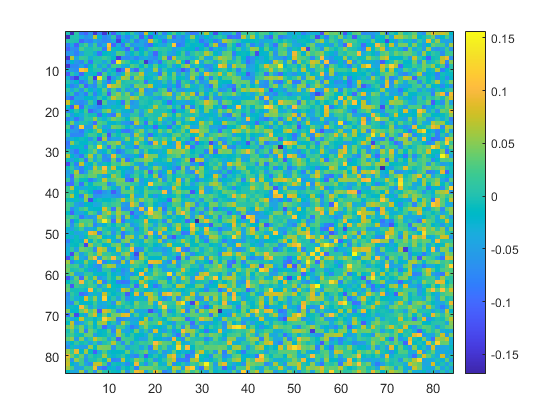}}
	%\vspace{-3.cm}
	\caption{{\bf Binarization of the Interaction Matrix:} The interaction matrix of the clique, ordered according to the species' abundance (with the most abundant species in the first row, and so on). In the upper panels, terms above the average ($\alpha_{i,j}>\alpha_R$) are colored yellow, while those below the average are blue (diagonal terms, which are all ones, are irrelevant). The lower panels display the actual values of $\alpha_{i,j}- \alpha_R$ for the same matrices, where weaker competition (negative values) is shown in blue and stronger competition in yellow. Results are presented for $S = 500$, $\sigma = 0.05$, and $\alpha_0 = 0.8$ (left), $0.4$ (middle), and $0.15$ (right). \label{fig5}}
\end{figure}

Since the elements of the interaction matrix are not all of the same magnitude, one has to specify which species compete strongly enough with a given species to be considered a competitor, and which do not.  Therefore, our metric for nestedness is based on relative competition strength. We take all the elements $\alpha_{i,j}$ in ${\cal Q}$ (excluding the diagonal elements) and subtract from them the average competition within the clique, $\alpha_R$. Now, any element greater than zero represents an interaction stronger than the average, while any element less than zero represents an interaction weaker than the average. We then assign a certain value (unity) to all elements greater than zero and a different value (zero) to elements less than zero. This binarization procedure transforms our interaction matrix into a matrix similar to the one described above. Figure \ref{fig5} illustrate this procedure for the interaction matrices of a few simulated cliques. 

Our nestedness metric is based on this binarization procedure. We calculate the Hamming distance between the binarized clique interaction matrix (neglecting the diagonal terms) and the ideal nested matrix in which the $\alpha_{i,j} = 0$  when $i+j \le Q+1$ and $\alpha_{i,j} = 1$ when $i+j > Q+1$, then divide the results by $Q(Q-1)$. The Hamming distance $d(\rm{clique},\rm{nested})$ is a number between zero (perfect nestedness) and $1/2$ (no nested structure). Our nestedness index is $1-d(\rm{clique},\rm{nested})$.  

Figure \ref{fig3c} shows that the symmetric cliques are nested in the $X<2$ region, and the smaller the value of $X$, the stronger the nestedness. In the asymmetric case, there is a much weaker, though still detectable, tendency toward being nested.

\begin{figure}
	%\vspace{-3.cm}
	\centering{
		\includegraphics[width=0.45\textwidth]{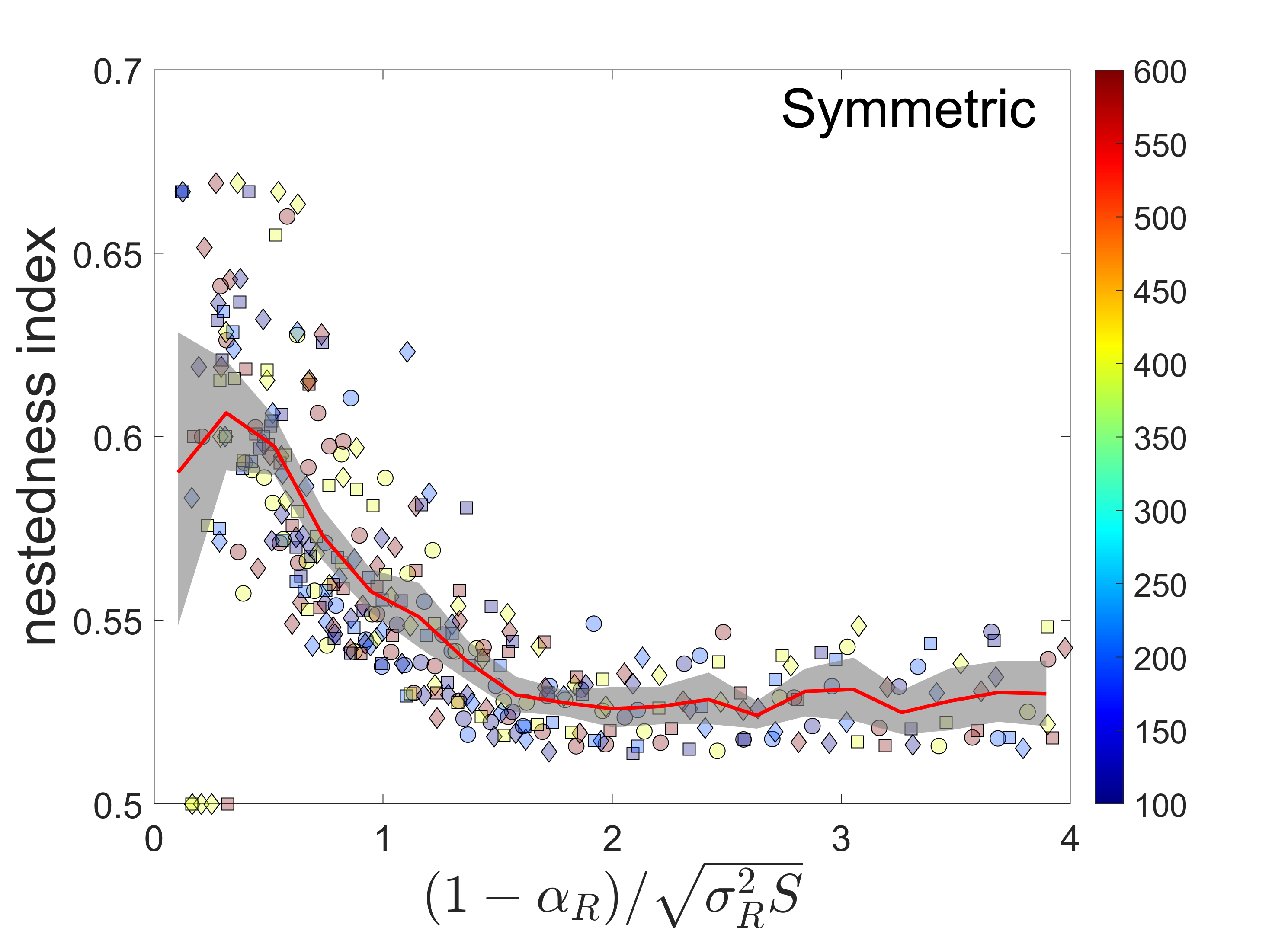}\includegraphics[width=0.45\textwidth]{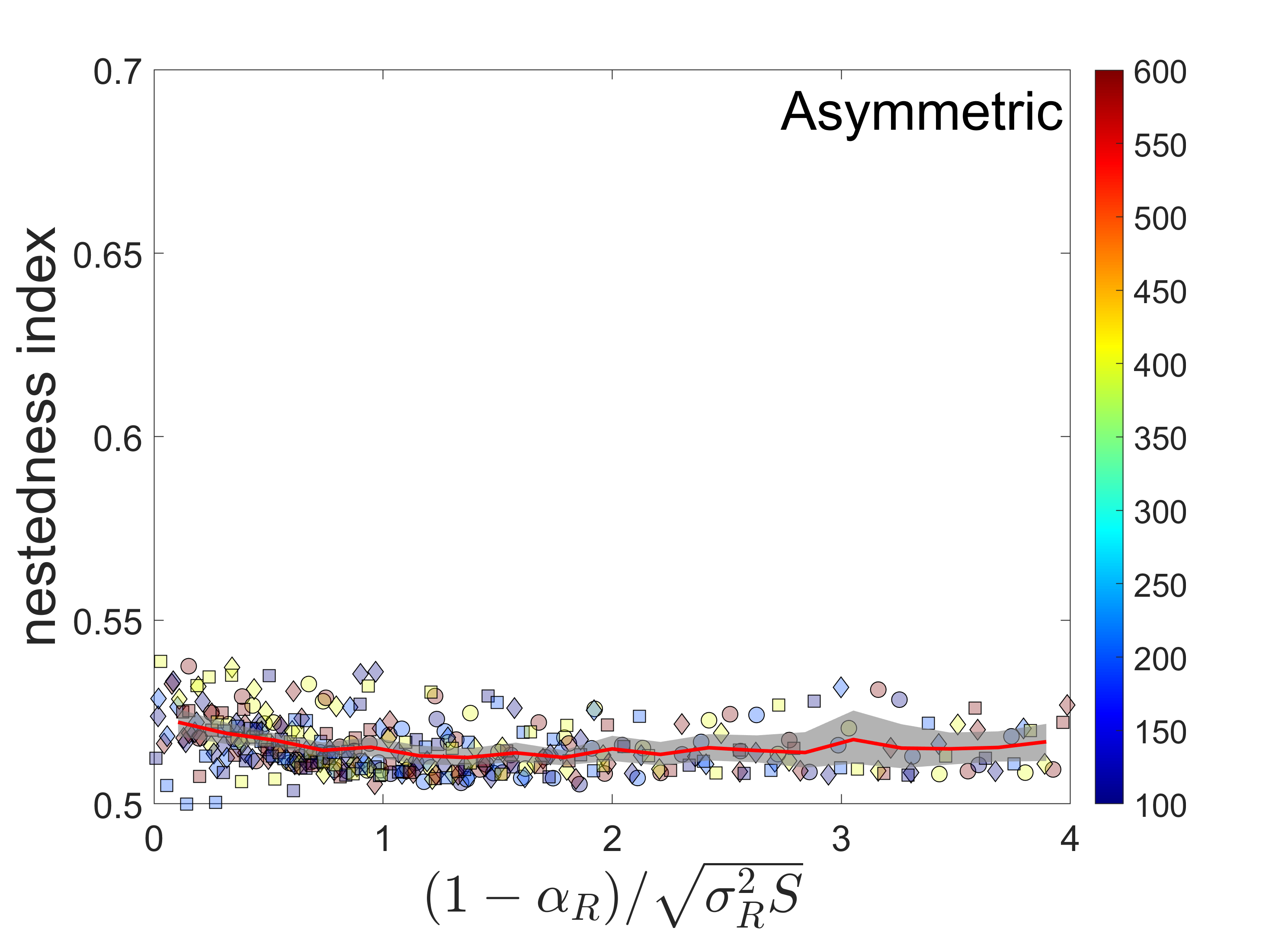} }
	%\vspace{-3.cm}
	\caption{{\bf Nestedness: }The nestedness index, defined as the Hamming distance between the binarized clique interaction matrix and a perfectly nested matrix, measures the degree of nested structure within a local community. A value of one indicates a perfectly nested community, while a value of zero suggests the absence of nestedness. Nestedness is more pronounced in symmetric cases and is much less pronounced, though still present, in asymmetric cases. The solid red line represents the moving average, the gray region shows the $95 \%$ confidence interval, and the other color and marker codes are identical to those in Fig \ref{fig3b}. \label{fig3c}}
\end{figure}

\subsection{Hyperuniformity}

In the theory of stochastic processes, there is a particular significance to matrices for which the sum of all rows (or columns) is equal. A similar phenomenon characterizes our problem: if the interaction matrix of the clique has constant row sums, for example,  
\[
{\cal Q} = \left( \renewcommand{\arraystretch}{1} % Adjust row height
\begin{array}{ccccc}
1 & 0.1 & 0.4 & 0.5 & 0 \\
0 & 1 & 0.3 & 0.1 & 0.6 \\
1 & 0 & 1 &  0 & 0 \\
0.7 & 0 & 0.3 & 1 & 0 \\
0.2 & 0 & 0 & 0.8 & 1
\end{array} \right)
\]
then the quantity 
\begin{equation}
W_i \equiv \sum_{j \neq i}^S \alpha_{i,j},
\end{equation}
 is independent of $i$. Without immigration, Eq. (\ref{eq1}) may be written, in that case, as,
\begin{equation}\label{eq10}
\frac{dN_i}{dt} = N_i \left( 1-N_i - W \sum_{j \neq i}^Q  N_j \right),
\end{equation}
and therefore 
\begin{equation}
N_i  = \frac{1}{1+W(Q-1)} \qquad \forall i
\end{equation}
is a stable solution.

\begin{figure}[h]
	%\vspace{-3.cm}
	\centering{
		 \includegraphics[width=8cm]{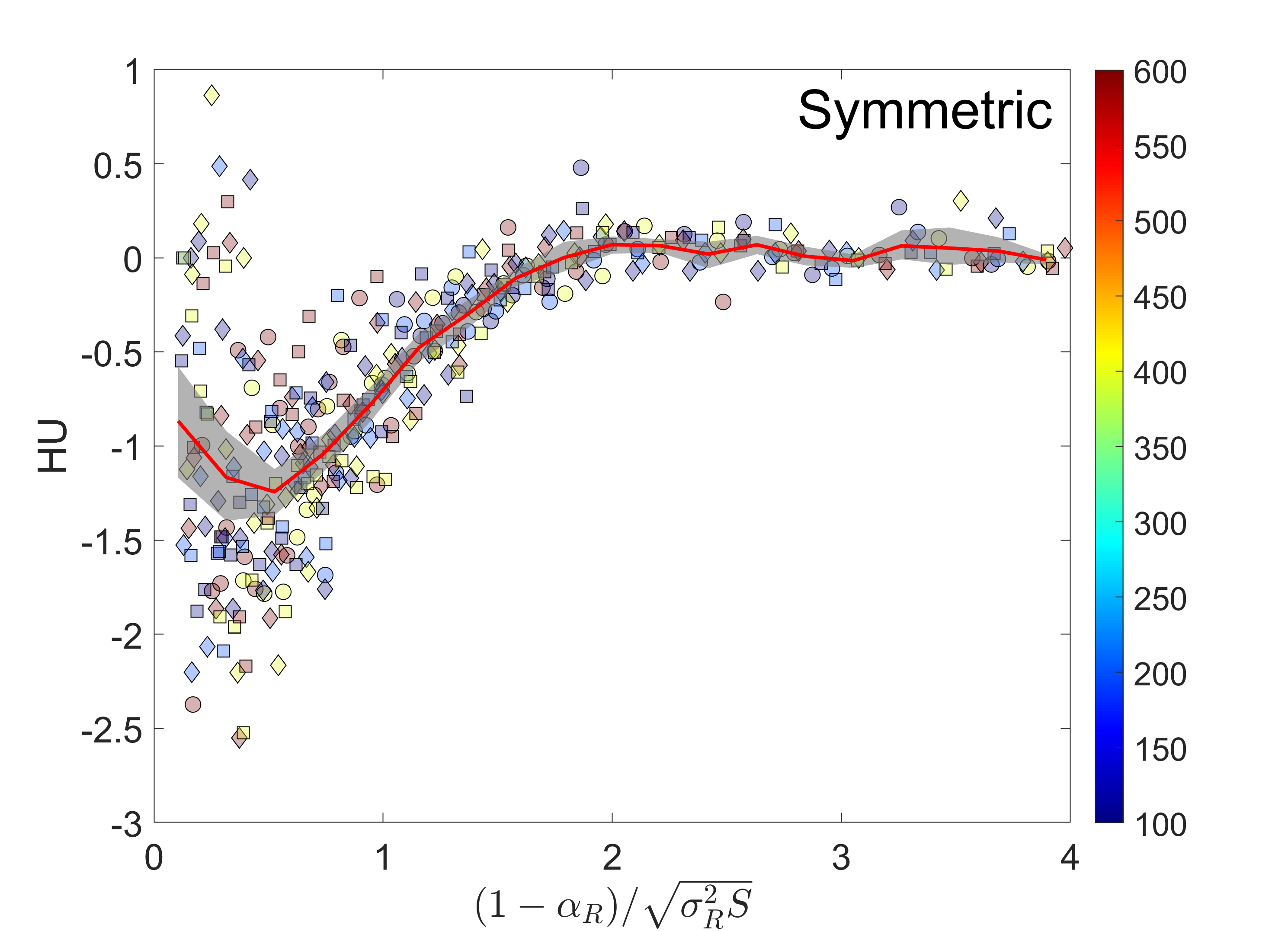} \includegraphics[width=8cm]{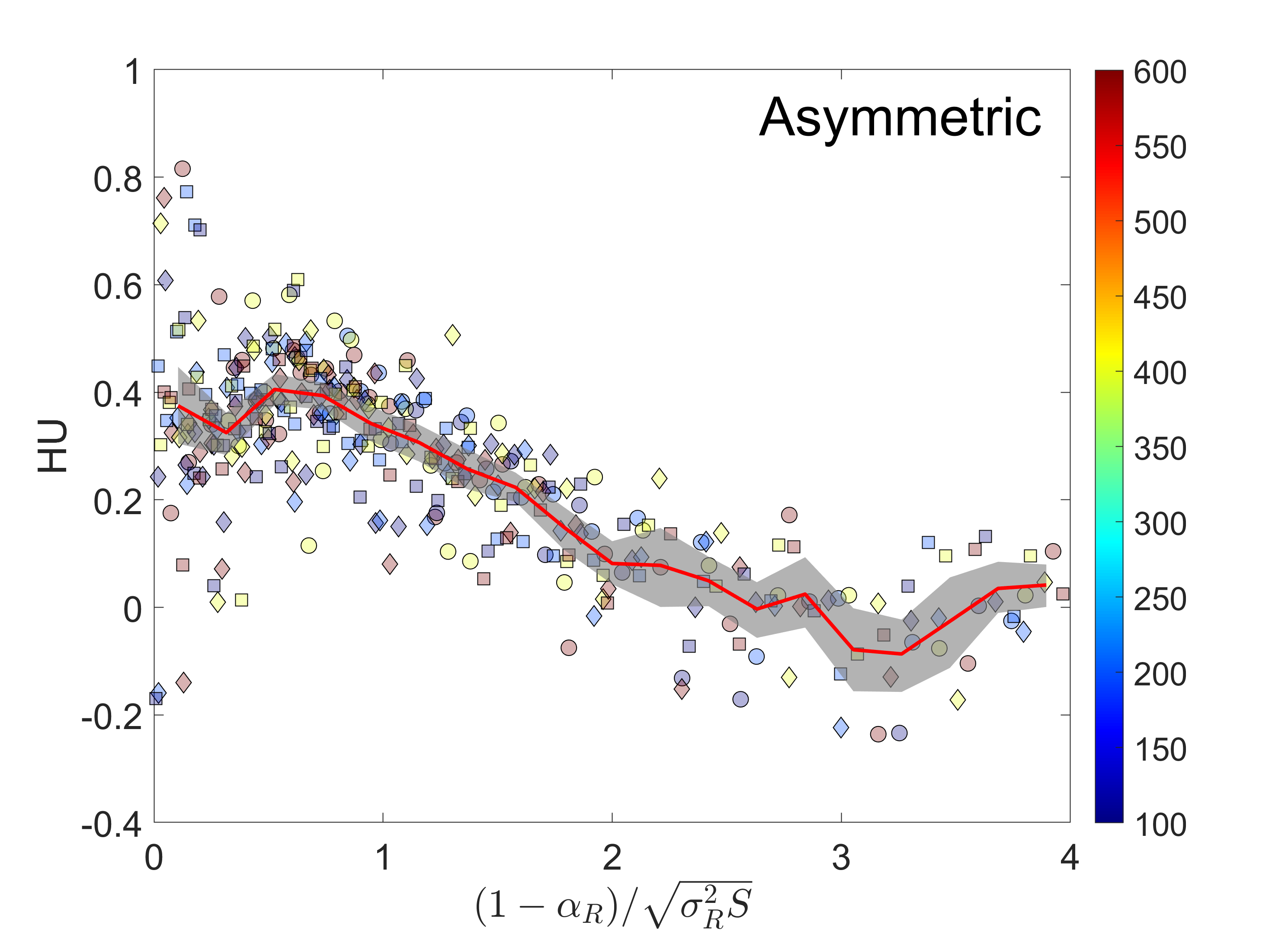}}
	%\vspace{-3.cm}
	\caption{{\bf Hyperuninformity: } $HU$ vs. $(1-\alpha_R)/\sigma_R \sqrt{N}$. When the interactions are asymmetric, the hyperuniformity is pronounced, whereas for symmetric matrices, the clique interaction matrix is less uniform than its randomly permuted counterpart. The solid red line represents the moving average, the gray region shows the $95 \%$ confidence interval, and the other color and marker codes are identical to those in Fig \ref{fig3b}.  \label{fig7}}
\end{figure}

In practice, the situation is, of course, less perfect. The $W_i$-s are not exactly the same, but their variations are less than what one would expect in a randomly assembled community. To quantify this, we compared the variance of the $W_i$s in the emergent clique community with the corresponding variance of the {\it same} interaction matrix when shuffled randomly. This provides us a hyperuniformity parameter, 
\begin{equation}
HU = 1- \frac{Var(W_{{\rm observed}})}{Var(W_{{\rm shuffled}})}.
\end{equation}
$HU = 0$ if there is no huperuniformity in the system,  $HU = 1$ when the observed interaction matrix is perfectly hyperuniform. $HU<0$ if the interaction matrix of the clique is less uniform than a random matrix. Figure \ref{fig7} shows that in the asymmetric case the trend towards hyperuniformity is pronounced, whereas in the symmetric case the trend is in the opposite direction, i.e., the clique matrix is subuniform rather than hyperuniform.
 
\section{Discussion} 

The mechanisms that govern the composition of a local community, given the species in the regional pool, have not yet been fully clarified. In particular, the species present in the local community may reflect dispersal limitations, environmental filtering, and the like. However, there is also interest in assembly processes that reflect interspecific competition, where the local clique is the result of constraints on the stability and feasibility of a group of competing species. In recent years, there has been immense interest in analyzing these systems~\cite{kessler2015generalized,bunin2017ecological,barbier2017cavity,barbier2021fingerprints,baron2023breakdown}.

If the interaction matrix is random, the primary factors determining the stability of the community are the strength of interspecific interaction and its variance. Therefore, we would expect the local community to be characterized by lower values of $\alpha$ and $\sigma$. As we have seen, the observed values of $\alpha_R$ and $\sigma_R$ in the local community are not low enough to account for stable coexistence of the clique. There are other characteristics, which are related to the structural properties of the clique interaction matrix. In this paper, we pointed out two such characteristics: the nested structure, mainly in symmetric systems, and the hyperuniformity in asymmetric systems.

There is still an important difference between the two cases. The reason why hyperuniformity facilitates coexistence is seemingly clear. As discussed above, in the limit of perfect hyperuniformity we obtain an equal abundance of all species, so small disturbances of this state will not lead to extinction. Extinction occurs when a specific species suffers significantly more than others, and reaching that point requires a strong distortion of the hyperuniform state. In contrast, it is not clear at all why a nested structure supports coexistence. There are studies on the importance of nestedness in bipartite systems~\cite{suweis2013emergence}, but that is not what we dealt with here.

While we may not fully understand how nestedness contributes to stability, we can trace the origin of this property: if two species within a clique have high abundance, it is likely that they exert less pressure on each other. This implies that, in the context of competitive interactions, the interaction matrix elements between them are smaller. On the other hand, hyperuniformity is a holistic property of the community as a whole, making it difficult to extract this characteristic solely from the analysis of two-species correlations.

We note that~\cite{barbier2021fingerprints} calculated how the value of the matrix element $\alpha_{i,j}$ depends on $N_i$ and $N_j$ for a given clique in the asymmetric case, and \cite{baron2023breakdown} extended these results to the general cases, including the symmetric one. In both of these papers, the results were obtained by conditioning the Gaussian ensemble of random matrices on the structure of a given clique. Here, on the other hand,  we implemented the opposite approach, calculating the clique structure numerically given the interaction matrix -- a technique that also allows us to satisfy the uninvasibility condition. The relationship between the structure identified here and the results of~\cite{baron2023breakdown} are not entirely clear to us and are under current investigation.

There are also other issues that still require clarification. First, one might ask whether the properties we have analyzed throughout this paper, namely nestedness and hyperuniformity, actually ensure stability. In other words, if we perform a random permutation of the elements of the ${\cal Q}$ matrix while preserving the desired structure of the interaction network, would that alone be sufficient to guarantee the coexistence of $Q$ species? Second, in our discussion, we focused on feasibility and stability, without examining the ability of the local community to resist invasion by additional species from the mainland. As discussed in~\cite{fried2017alternative},
 the property of invadability significantly constrains the space of permissible island communities, at least within the schematic models explored there. The importance of invadability to the network's structure is a subject that requires further investigation. These open questions could provide a basis for additional research on the problem at hand, a problem whose critical importance becomes clearer as more data on biological communities across various domains is collected.

\bibliography{ref}

\begin{thebibliography}{29}
\expandafter\ifx\csname natexlab\endcsname\relax\def\natexlab#1{#1}\fi
\expandafter\ifx\csname bibnamefont\endcsname\relax
  \def\bibnamefont#1{#1}\fi
\expandafter\ifx\csname bibfnamefont\endcsname\relax
  \def\bibfnamefont#1{#1}\fi
\expandafter\ifx\csname citenamefont\endcsname\relax
  \def\citenamefont#1{#1}\fi
\expandafter\ifx\csname url\endcsname\relax
  \def\url#1{\texttt{#1}}\fi
\expandafter\ifx\csname urlprefix\endcsname\relax\def\urlprefix{URL }\fi
\providecommand{\bibinfo}[2]{#2}
\providecommand{\eprint}[2][]{\url{#2}}

\bibitem[{\citenamefont{Chesson}(2000)}]{chesson2000mechanisms}
\bibinfo{author}{\bibfnamefont{P.}~\bibnamefont{Chesson}}, \bibinfo{journal}{Annual Review of Ecology and Systematics} \textbf{\bibinfo{volume}{31}}, \bibinfo{pages}{343} (\bibinfo{year}{2000}).

\bibitem[{\citenamefont{Levine et~al.}(2017)\citenamefont{Levine, Bascompte, Adler, and Allesina}}]{levine2017beyond}
\bibinfo{author}{\bibfnamefont{J.~M.} \bibnamefont{Levine}}, \bibinfo{author}{\bibfnamefont{J.}~\bibnamefont{Bascompte}}, \bibinfo{author}{\bibfnamefont{P.~B.} \bibnamefont{Adler}}, \bibnamefont{and} \bibinfo{author}{\bibfnamefont{S.}~\bibnamefont{Allesina}}, \bibinfo{journal}{Nature} \textbf{\bibinfo{volume}{546}}, \bibinfo{pages}{56} (\bibinfo{year}{2017}).

\bibitem[{\citenamefont{Ter~Steege et~al.}(2013)\citenamefont{Ter~Steege, Pitman, Sabatier, Baraloto, Salom{\~a}o, Guevara, Phillips, Castilho, Magnusson, Molino et~al.}}]{ter2013hyperdominance}
\bibinfo{author}{\bibfnamefont{H.}~\bibnamefont{Ter~Steege}}, \bibinfo{author}{\bibfnamefont{N.~C.} \bibnamefont{Pitman}}, \bibinfo{author}{\bibfnamefont{D.}~\bibnamefont{Sabatier}}, \bibinfo{author}{\bibfnamefont{C.}~\bibnamefont{Baraloto}}, \bibinfo{author}{\bibfnamefont{R.~P.} \bibnamefont{Salom{\~a}o}}, \bibinfo{author}{\bibfnamefont{J.~E.} \bibnamefont{Guevara}}, \bibinfo{author}{\bibfnamefont{O.~L.} \bibnamefont{Phillips}}, \bibinfo{author}{\bibfnamefont{C.~V.} \bibnamefont{Castilho}}, \bibinfo{author}{\bibfnamefont{W.~E.} \bibnamefont{Magnusson}}, \bibinfo{author}{\bibfnamefont{J.-F.} \bibnamefont{Molino}}, \bibnamefont{et~al.}, \bibinfo{journal}{Science} \textbf{\bibinfo{volume}{342}}, \bibinfo{pages}{1243092} (\bibinfo{year}{2013}).

\bibitem[{\citenamefont{Volkov et~al.}(2003)\citenamefont{Volkov, Banavar, Hubbell, and Maritan}}]{maritan1}
\bibinfo{author}{\bibfnamefont{I.}~\bibnamefont{Volkov}}, \bibinfo{author}{\bibfnamefont{J.~R.} \bibnamefont{Banavar}}, \bibinfo{author}{\bibfnamefont{S.~P.} \bibnamefont{Hubbell}}, \bibnamefont{and} \bibinfo{author}{\bibfnamefont{A.}~\bibnamefont{Maritan}}, \bibinfo{journal}{Nature} \textbf{\bibinfo{volume}{424}}, \bibinfo{pages}{1035} (\bibinfo{year}{2003}).

\bibitem[{\citenamefont{Connolly et~al.}(2014)\citenamefont{Connolly, MacNeil, Caley, Knowlton, Cripps, Hisano, Thibaut, Bhattacharya, Benedetti-Cecchi, Brainard et~al.}}]{connolly2014commonness}
\bibinfo{author}{\bibfnamefont{S.~R.} \bibnamefont{Connolly}}, \bibinfo{author}{\bibfnamefont{M.~A.} \bibnamefont{MacNeil}}, \bibinfo{author}{\bibfnamefont{M.~J.} \bibnamefont{Caley}}, \bibinfo{author}{\bibfnamefont{N.}~\bibnamefont{Knowlton}}, \bibinfo{author}{\bibfnamefont{E.}~\bibnamefont{Cripps}}, \bibinfo{author}{\bibfnamefont{M.}~\bibnamefont{Hisano}}, \bibinfo{author}{\bibfnamefont{L.~M.} \bibnamefont{Thibaut}}, \bibinfo{author}{\bibfnamefont{B.~D.} \bibnamefont{Bhattacharya}}, \bibinfo{author}{\bibfnamefont{L.}~\bibnamefont{Benedetti-Cecchi}}, \bibinfo{author}{\bibfnamefont{R.~E.} \bibnamefont{Brainard}}, \bibnamefont{et~al.}, \bibinfo{journal}{Proceedings of the National Academy of Sciences} \textbf{\bibinfo{volume}{111}}, \bibinfo{pages}{8524} (\bibinfo{year}{2014}).

\bibitem[{\citenamefont{Stomp et~al.}(2011)\citenamefont{Stomp, Huisman, Mittelbach, Litchman, and Klausmeier}}]{stomp2011large}
\bibinfo{author}{\bibfnamefont{M.}~\bibnamefont{Stomp}}, \bibinfo{author}{\bibfnamefont{J.}~\bibnamefont{Huisman}}, \bibinfo{author}{\bibfnamefont{G.~G.} \bibnamefont{Mittelbach}}, \bibinfo{author}{\bibfnamefont{E.}~\bibnamefont{Litchman}}, \bibnamefont{and} \bibinfo{author}{\bibfnamefont{C.~A.} \bibnamefont{Klausmeier}}, \bibinfo{journal}{Ecology} \textbf{\bibinfo{volume}{92}}, \bibinfo{pages}{2096} (\bibinfo{year}{2011}).

\bibitem[{\citenamefont{Friedman et~al.}(2017)\citenamefont{Friedman, Higgins, and Gore}}]{friedman2017community}
\bibinfo{author}{\bibfnamefont{J.}~\bibnamefont{Friedman}}, \bibinfo{author}{\bibfnamefont{L.~M.} \bibnamefont{Higgins}}, \bibnamefont{and} \bibinfo{author}{\bibfnamefont{J.}~\bibnamefont{Gore}}, \bibinfo{journal}{Nature ecology \& evolution} \textbf{\bibinfo{volume}{1}}, \bibinfo{pages}{1} (\bibinfo{year}{2017}).

\bibitem[{\citenamefont{Bashan et~al.}(2016)\citenamefont{Bashan, Gibson, Friedman, Carey, Weiss, Hohmann, and Liu}}]{bashan2016universality}
\bibinfo{author}{\bibfnamefont{A.}~\bibnamefont{Bashan}}, \bibinfo{author}{\bibfnamefont{T.~E.} \bibnamefont{Gibson}}, \bibinfo{author}{\bibfnamefont{J.}~\bibnamefont{Friedman}}, \bibinfo{author}{\bibfnamefont{V.~J.} \bibnamefont{Carey}}, \bibinfo{author}{\bibfnamefont{S.~T.} \bibnamefont{Weiss}}, \bibinfo{author}{\bibfnamefont{E.~L.} \bibnamefont{Hohmann}}, \bibnamefont{and} \bibinfo{author}{\bibfnamefont{Y.-Y.} \bibnamefont{Liu}}, \bibinfo{journal}{Nature} \textbf{\bibinfo{volume}{534}}, \bibinfo{pages}{259} (\bibinfo{year}{2016}).

\bibitem[{\citenamefont{Yonatan et~al.}(2022)\citenamefont{Yonatan, Amit, Friedman, and Bashan}}]{yonatan2022complexity}
\bibinfo{author}{\bibfnamefont{Y.}~\bibnamefont{Yonatan}}, \bibinfo{author}{\bibfnamefont{G.}~\bibnamefont{Amit}}, \bibinfo{author}{\bibfnamefont{J.}~\bibnamefont{Friedman}}, \bibnamefont{and} \bibinfo{author}{\bibfnamefont{A.}~\bibnamefont{Bashan}}, \bibinfo{journal}{Nature Ecology \& Evolution} pp. \bibinfo{pages}{1--8} (\bibinfo{year}{2022}).

\bibitem[{\citenamefont{Fisher and Mehta}(2013)}]{fisher2013niche}
\bibinfo{author}{\bibfnamefont{C.~K.} \bibnamefont{Fisher}} \bibnamefont{and} \bibinfo{author}{\bibfnamefont{P.}~\bibnamefont{Mehta}}, \bibinfo{journal}{arXiv preprint arXiv:1308.2969}  (\bibinfo{year}{2013}).

\bibitem[{\citenamefont{Kessler and Shnerb}(2015)}]{kessler2015generalized}
\bibinfo{author}{\bibfnamefont{D.~A.} \bibnamefont{Kessler}} \bibnamefont{and} \bibinfo{author}{\bibfnamefont{N.~M.} \bibnamefont{Shnerb}}, \bibinfo{journal}{Physical Review E} \textbf{\bibinfo{volume}{91}}, \bibinfo{pages}{042705} (\bibinfo{year}{2015}).

\bibitem[{\citenamefont{Bunin}(2017)}]{bunin2017ecological}
\bibinfo{author}{\bibfnamefont{G.}~\bibnamefont{Bunin}}, \bibinfo{journal}{Physical Review E} \textbf{\bibinfo{volume}{95}}, \bibinfo{pages}{042414} (\bibinfo{year}{2017}).

\bibitem[{\citenamefont{Barbier et~al.}(2018)\citenamefont{Barbier, Arnoldi, Bunin, and Loreau}}]{barbier2018generic}
\bibinfo{author}{\bibfnamefont{M.}~\bibnamefont{Barbier}}, \bibinfo{author}{\bibfnamefont{J.-F.} \bibnamefont{Arnoldi}}, \bibinfo{author}{\bibfnamefont{G.}~\bibnamefont{Bunin}}, \bibnamefont{and} \bibinfo{author}{\bibfnamefont{M.}~\bibnamefont{Loreau}}, \bibinfo{journal}{Proceedings of the National Academy of Sciences} \textbf{\bibinfo{volume}{115}}, \bibinfo{pages}{2156} (\bibinfo{year}{2018}).

\bibitem[{\citenamefont{Fried et~al.}(2016)\citenamefont{Fried, Kessler, and Shnerb}}]{fried2016communities}
\bibinfo{author}{\bibfnamefont{Y.}~\bibnamefont{Fried}}, \bibinfo{author}{\bibfnamefont{D.~A.} \bibnamefont{Kessler}}, \bibnamefont{and} \bibinfo{author}{\bibfnamefont{N.~M.} \bibnamefont{Shnerb}}, \bibinfo{journal}{Scientific reports} \textbf{\bibinfo{volume}{6}}, \bibinfo{pages}{35648} (\bibinfo{year}{2016}).

\bibitem[{\citenamefont{Fried et~al.}(2017)\citenamefont{Fried, Shnerb, and Kessler}}]{fried2017alternative}
\bibinfo{author}{\bibfnamefont{Y.}~\bibnamefont{Fried}}, \bibinfo{author}{\bibfnamefont{N.~M.} \bibnamefont{Shnerb}}, \bibnamefont{and} \bibinfo{author}{\bibfnamefont{D.~A.} \bibnamefont{Kessler}}, \bibinfo{journal}{Physical Review E} \textbf{\bibinfo{volume}{96}}, \bibinfo{pages}{012412} (\bibinfo{year}{2017}).

\bibitem[{\citenamefont{Arnoulx~de Pirey and Bunin}(2024)}]{arnoulx2024many}
\bibinfo{author}{\bibfnamefont{T.}~\bibnamefont{Arnoulx~de Pirey}} \bibnamefont{and} \bibinfo{author}{\bibfnamefont{G.}~\bibnamefont{Bunin}}, \bibinfo{journal}{Physical Review X} \textbf{\bibinfo{volume}{14}}, \bibinfo{pages}{011037} (\bibinfo{year}{2024}).

\bibitem[{\citenamefont{Mallmin et~al.}(2024)\citenamefont{Mallmin, Traulsen, and De~Monte}}]{mallmin2024chaotic}
\bibinfo{author}{\bibfnamefont{E.}~\bibnamefont{Mallmin}}, \bibinfo{author}{\bibfnamefont{A.}~\bibnamefont{Traulsen}}, \bibnamefont{and} \bibinfo{author}{\bibfnamefont{S.}~\bibnamefont{De~Monte}}, \bibinfo{journal}{Proceedings of the National Academy of Sciences} \textbf{\bibinfo{volume}{121}}, \bibinfo{pages}{e2312822121} (\bibinfo{year}{2024}).

\bibitem[{\citenamefont{Tilman}(1982)}]{tilman1982resource}
\bibinfo{author}{\bibfnamefont{D.}~\bibnamefont{Tilman}}, \emph{\bibinfo{title}{Resource competition and community structure}}, \bibinfo{number}{17} (\bibinfo{publisher}{Princeton university press}, \bibinfo{year}{1982}).

\bibitem[{\citenamefont{May}(1972)}]{may1972will}
\bibinfo{author}{\bibfnamefont{R.~M.} \bibnamefont{May}}, \bibinfo{journal}{Nature} \textbf{\bibinfo{volume}{238}}, \bibinfo{pages}{413} (\bibinfo{year}{1972}).

\bibitem[{\citenamefont{Song et~al.}(2019)\citenamefont{Song, Barab{\'a}s, and Saavedra}}]{song2019consequences}
\bibinfo{author}{\bibfnamefont{C.}~\bibnamefont{Song}}, \bibinfo{author}{\bibfnamefont{G.}~\bibnamefont{Barab{\'a}s}}, \bibnamefont{and} \bibinfo{author}{\bibfnamefont{S.}~\bibnamefont{Saavedra}}, \bibinfo{journal}{The American Naturalist} \textbf{\bibinfo{volume}{194}}, \bibinfo{pages}{627} (\bibinfo{year}{2019}).

\bibitem[{\citenamefont{Allesina and Tang}(2015)}]{allesina2015stability}
\bibinfo{author}{\bibfnamefont{S.}~\bibnamefont{Allesina}} \bibnamefont{and} \bibinfo{author}{\bibfnamefont{S.}~\bibnamefont{Tang}}, \bibinfo{journal}{Population Ecology} \textbf{\bibinfo{volume}{57}}, \bibinfo{pages}{63} (\bibinfo{year}{2015}).

\bibitem[{\citenamefont{Allesina and Tang}(2012)}]{allesina2012stability}
\bibinfo{author}{\bibfnamefont{S.}~\bibnamefont{Allesina}} \bibnamefont{and} \bibinfo{author}{\bibfnamefont{S.}~\bibnamefont{Tang}}, \bibinfo{journal}{Nature} \textbf{\bibinfo{volume}{483}}, \bibinfo{pages}{205} (\bibinfo{year}{2012}).

\bibitem[{\citenamefont{Bizeul and Najim}(2021)}]{bizeul2021positive}
\bibinfo{author}{\bibfnamefont{P.}~\bibnamefont{Bizeul}} \bibnamefont{and} \bibinfo{author}{\bibfnamefont{J.}~\bibnamefont{Najim}}, \bibinfo{journal}{Proceedings of the American Mathematical Society} \textbf{\bibinfo{volume}{149}}, \bibinfo{pages}{2333} (\bibinfo{year}{2021}).

\bibitem[{\citenamefont{Hu et~al.}(2022)\citenamefont{Hu, Amor, Barbier, Bunin, and Gore}}]{hu2022emergent}
\bibinfo{author}{\bibfnamefont{J.}~\bibnamefont{Hu}}, \bibinfo{author}{\bibfnamefont{D.~R.} \bibnamefont{Amor}}, \bibinfo{author}{\bibfnamefont{M.}~\bibnamefont{Barbier}}, \bibinfo{author}{\bibfnamefont{G.}~\bibnamefont{Bunin}}, \bibnamefont{and} \bibinfo{author}{\bibfnamefont{J.}~\bibnamefont{Gore}}, \bibinfo{journal}{Science} \textbf{\bibinfo{volume}{378}}, \bibinfo{pages}{85} (\bibinfo{year}{2022}).

\bibitem[{\citenamefont{Beckett et~al.}(2014)\citenamefont{Beckett, Boulton, and Williams}}]{beckett2014falcon}
\bibinfo{author}{\bibfnamefont{S.~J.} \bibnamefont{Beckett}}, \bibinfo{author}{\bibfnamefont{C.~A.} \bibnamefont{Boulton}}, \bibnamefont{and} \bibinfo{author}{\bibfnamefont{H.~T.} \bibnamefont{Williams}}, \bibinfo{journal}{F1000Research} \textbf{\bibinfo{volume}{3}} (\bibinfo{year}{2014}).

\bibitem[{\citenamefont{Barbier and Arnoldi}(2017)}]{barbier2017cavity}
\bibinfo{author}{\bibfnamefont{M.}~\bibnamefont{Barbier}} \bibnamefont{and} \bibinfo{author}{\bibfnamefont{J.-F.} \bibnamefont{Arnoldi}}, \bibinfo{journal}{bioRxiv} p. \bibinfo{pages}{147728} (\bibinfo{year}{2017}).

\bibitem[{\citenamefont{Barbier et~al.}(2021)\citenamefont{Barbier, de~Mazancourt, Loreau, and Bunin}}]{barbier2021fingerprints}
\bibinfo{author}{\bibfnamefont{M.}~\bibnamefont{Barbier}}, \bibinfo{author}{\bibfnamefont{C.}~\bibnamefont{de~Mazancourt}}, \bibinfo{author}{\bibfnamefont{M.}~\bibnamefont{Loreau}}, \bibnamefont{and} \bibinfo{author}{\bibfnamefont{G.}~\bibnamefont{Bunin}}, \bibinfo{journal}{Physical Review X} \textbf{\bibinfo{volume}{11}}, \bibinfo{pages}{011009} (\bibinfo{year}{2021}).

\bibitem[{\citenamefont{Baron et~al.}(2023)\citenamefont{Baron, Jewell, Ryder, and Galla}}]{baron2023breakdown}
\bibinfo{author}{\bibfnamefont{J.~W.} \bibnamefont{Baron}}, \bibinfo{author}{\bibfnamefont{T.~J.} \bibnamefont{Jewell}}, \bibinfo{author}{\bibfnamefont{C.}~\bibnamefont{Ryder}}, \bibnamefont{and} \bibinfo{author}{\bibfnamefont{T.}~\bibnamefont{Galla}}, \bibinfo{journal}{Physical Review Letters} \textbf{\bibinfo{volume}{130}}, \bibinfo{pages}{137401} (\bibinfo{year}{2023}).

\bibitem[{\citenamefont{Suweis et~al.}(2013)\citenamefont{Suweis, Simini, Banavar, and Maritan}}]{suweis2013emergence}
\bibinfo{author}{\bibfnamefont{S.}~\bibnamefont{Suweis}}, \bibinfo{author}{\bibfnamefont{F.}~\bibnamefont{Simini}}, \bibinfo{author}{\bibfnamefont{J.~R.} \bibnamefont{Banavar}}, \bibnamefont{and} \bibinfo{author}{\bibfnamefont{A.}~\bibnamefont{Maritan}}, \bibinfo{journal}{Nature} \textbf{\bibinfo{volume}{500}}, \bibinfo{pages}{449} (\bibinfo{year}{2013}).

\end{thebibliography}

\clearpage
\appendix

\section{Assessing stability of a clique} \label{apA}

Throughout this paper, we have discussed the "clique,'' which is the group of species on the island with an abundance greater than the square root of the immigration rate $\lambda$. In some cases, the clique is stable and uninvadable, as described in \ref{fig1}. In other cases, this local community is unstable, either because the species within the clique cannot coexist (so if we leave them for an extended period, some will go extinct) or because species not included in the clique can invade from the regional community, as seen, for example, in the right panel of Fig. \ref{fig3}.

In most cases we dealt with, it was not important to distinguish between stable cliques and those that are not stable. Even in cases where the cliques are unstable, they still reflect the current state of the community on the island. Since we deal with generic dynamics, examining the structural properties of the interaction network for the typical community is just as interesting as the corresponding examination for the stable state. However, at several points throughout the paper, particularly in the discussion of what confers stability to a clique (Section \ref{sec3}), we needed to distinguish between stable and uninvadable states and those that are not. Here, we will explain how we distinguished between the two types of cliques in the numerical experiments we conducted.

We examined the stability of a clique by comparing the observed abundance vector (i.e., the one obtained from the simulation at a given moment) with the abundance vector obtained from solving the linear problem 
\begin{equation} \label{apeq1}
{\cal Q} {\cal N} = \mathbb{1},
\end{equation}  
where ${\cal Q}$ is the interaction matrix of the clique. If the clique is stable, the result should be identical up to numerical truncation errors and the effect of the immigration rate $\lambda$. If the clique is not stable, either because it is a chaotic system or because it is susceptible to invasion, the abundances of the species will change constantly, and therefore the error will be larger. Theoretically, we could encounter a solution to the linear problem that is feasible but not stable, but there is no reason for the system to reach an unstable fixed point through numerical integration. Therefore, it is reasonable to assume that if the solution to the linear problem equals the abundances obtained dynamically, the system is stable.

Our stability parameter is thus, 
\begin{equation}
    St = \sqrt{\sum_{i=1..Q} \left( {\cal N}_i - N_i \right)^2},
\end{equation}
where $N_i$ are the numerically calculated abundances and ${\cal N}_i$ are the predicted values according to Eq. (\ref{apeq1}).

\begin{figure}[h]
	%\vspace{-3.cm}
	\centering{
		 \includegraphics[width=8cm]{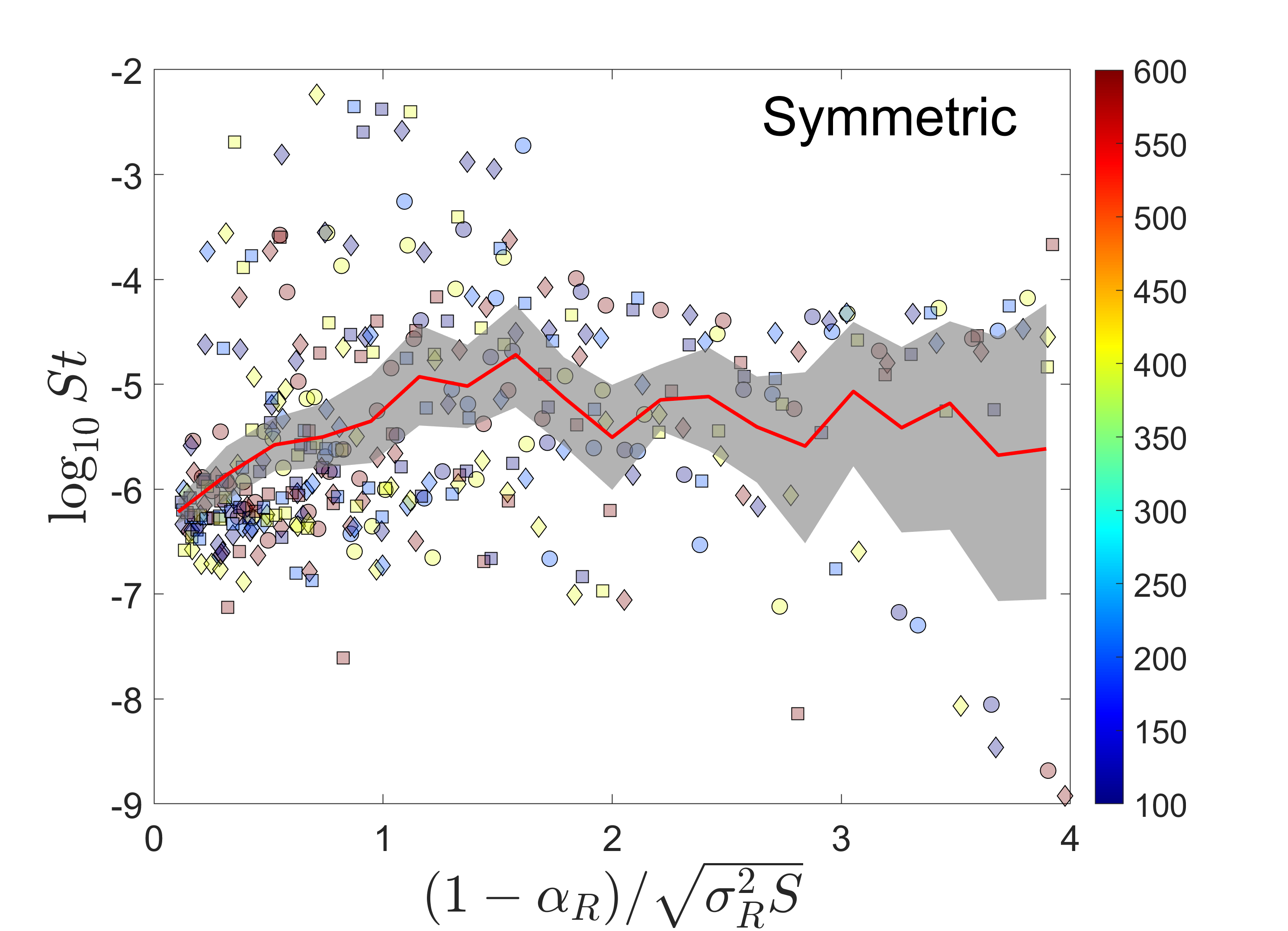} \includegraphics[width=8cm]{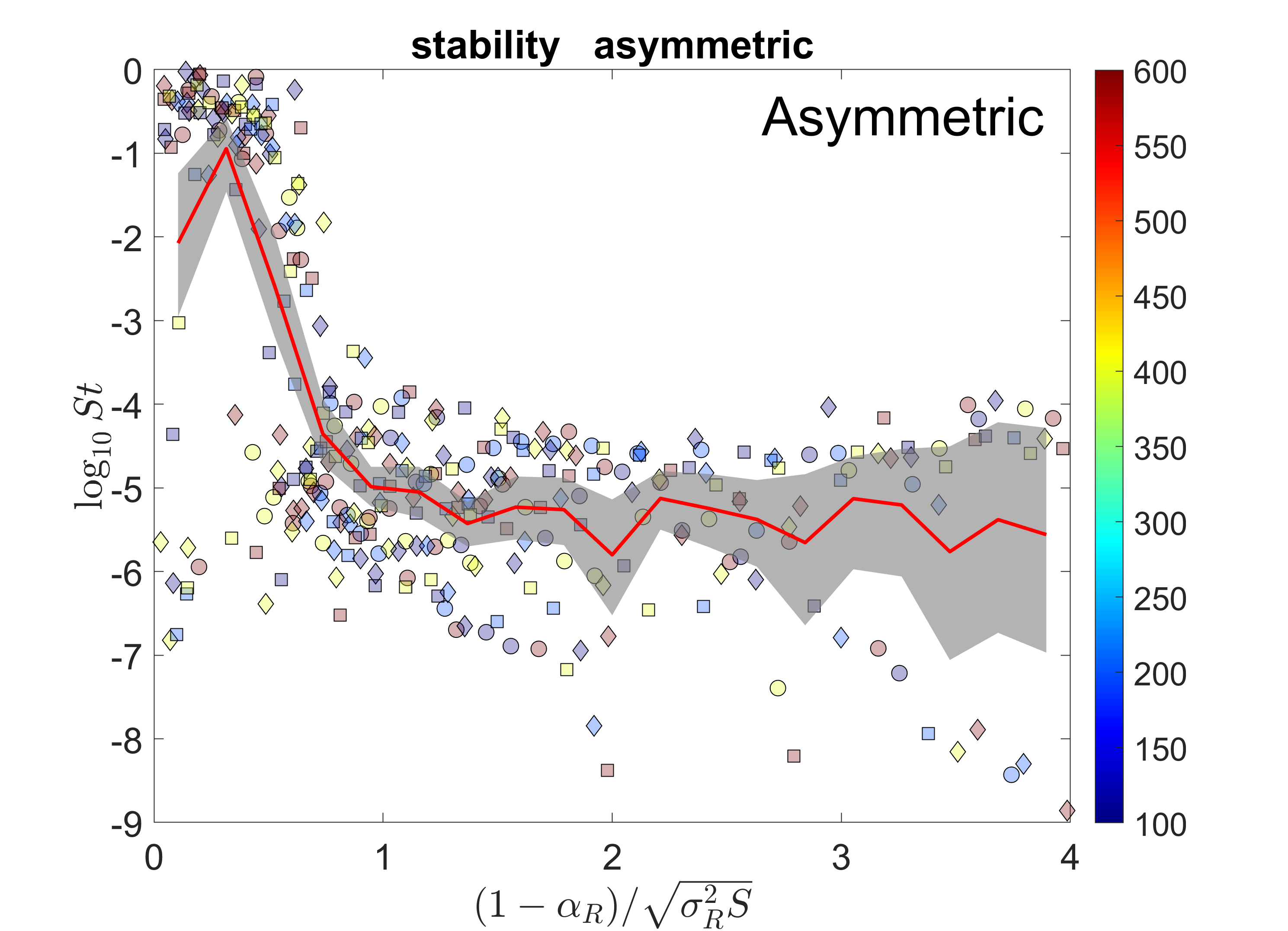}}
	%\vspace{-3.cm}
	\caption{{\bf Clique stability: } The stability index, $St$, vs. $(1-\alpha_R)/\sigma_R \sqrt{N}$. When the interactions are asymmetric, most of the observed cliques are stable, whereas instability is the common case for asymmetric systems with $X<1$.  \label{apfig2}}
\end{figure}

Figure \ref{apfig2} shows the results for the symmetric and the asymmetric cases. To be on the safe side, we took $St=10^{-4}$ as the critical value above which the clique is considered unstable. Thus, only these cliques were examined for the feasibility of permutation in Fig. \ref{fig3c} of the main text. 

It can be seen that for symmetric systems, the clique is generally stable, with a small number of exceptions reflecting long convergence times of the system. In contrast, in asymmetric systems below $X = 1$, the generic case is one of an unstable clique, due to the transition to periodic or chaotic dynamics.

\end{document}